\documentclass{aastex}
\usepackage{spr-astr-addons}
\usepackage{url}
\usepackage{graphicx,epsfig}
\usepackage{amsmath}
\RequirePackage{color}

\def\spose#1{\hbox to 0pt{#1\hss}}
\def\lta{\mathrel{\spose{\lower 3pt\hbox{$\mathchar"218$}}
     \raise 2.0pt\hbox{$\mathchar"13C$}}}
\def\gta{\mathrel{\spose{\lower 3pt\hbox{$\mathchar"218$}}
     \raise 2.0pt\hbox{$\mathchar"13E$}}}

\begin{document}

\setlength{\overfullrule}{0pt}

\title{
Dead zones of classical habitability \\
in stellar binary systems
}

\shorttitle{Dead zones in stellar binary systems}
\shortauthors{Moorman et al.}

\author{S. Y. Moorman\altaffilmark{1}} \author{Zh. Wang\altaffilmark{1}} \and \author{M. Cuntz\altaffilmark{1}}

\altaffiltext{1}{Department of Physics, University of Texas at Arlington, Box 19059,
       Arlington, TX 76019, USA.\\
       email: sarah.moorman@mavs.uta.edu \\
       email: zhaopeng.wang@mavs.uta.edu \\
       email: cuntz@uta.edu
}

\begin{abstract}
   Although habitability, defined as the general possibility of hosting life,
   is expected to occur under a broad range of conditions, the standard scenario
   to allow for habitable environments is often described through habitable zones
   (HZs).  Previous work indicates that stellar binary systems typically possess
   S-type or P-type HZs, with the S-type HZs forming ring-type structures around
   the individual stars and P-type HZs forming similar structures around both stars,
   if considered a pair.  However, depending on the stellar and orbital parameters
   of the system, typically, there are also regions within the systems outside of the
   HZs, referred to as dead zones (DZs).  In this study, we will convey quantitative
   information on the width and location of DZs for various systems.  The results
   will also depend on the definition of the stellar HZs as those are informed by
   the planetary climate models.
\end{abstract}


\section{Introduction}

For more than a decade, the concept of habitable planets in stellar binary 
systems has been a topic of great interest, most notably due to the numerous 
discoveries of planets within those systems in addition to the large percentage 
of stellar binary and higher order systems observed in our galactic neighborhood 
\citep{duq91,pat02,egg04,rag06,rag10,roe12}.  Such systems provide a new and unique 
set of potential candidates to be explored in the search for life beyond Earth; however, 
certain candidate locations appear to be statistically preferred over others as
these statistical studies underlie the notion of habitability.

Habitability is just as it implies, the capability to be habitable, and it carries with it a list 
of requirements, some of which could be argued as classical whereas others exotic.  
One traditional criterion for a target location to be habitable is that it must reside within 
a physically defined region called the habitable zone.  Habitable zones (HZs) are classically 
defined as spherical regions around a star or system of stars in which terrestrial planets 
could potentially possess surface temperatures allowing the existence of liquid water, given a 
sufficiently dense atmosphere \citep[e.g.,][ and related work]{kas93}.  The classical HZ is characterized 
by the greenhouse effect of CO$_\mathrm{2}$ and H$_\mathrm{2}$O vapor with the accompanying 
assumption that rocky planets are geologically active, allowing them to regulate their
CO$_\mathrm{2}$ in their N$_\mathrm{2}$--CO$_\mathrm{2}$--H$_\mathrm{2}$O 
atmospheres.

However, the existence of HZs does not ultimately signify 
the possibility of life as the latter is expected to depend on additional factors; e.g.,
\cite{des08}, and references therein.  Conversely, life may even possibly be based on non-standard
biochemistries \citep[e.g.,][]{bai04}.  Nonetheless, the standard definition of HZs assumes
regions around stars with the general possibility for liquid water on possible system planets
(or moons); however, following previous analyses by, e.g., \cite{kas03}, \cite{lam09}, and
\cite{coc16}, the notion of circumstellar habitability is significantly more complex.
Although, even if the water-based definition is adopted, as customarily done, the imposed
requirements do not  necessarily encompass all instances of potential habitability.  For
example, planets on eccentric orbits could potentially remain habitable (i.e., retain enough heat
to allow for surface water) even with portions of their orbits located outside of the classical HZ 
\citep{wil02}.  Moreover, large moons in orbit about the giant planets of our Solar System
currently residing beyond the traditionally defined HZ could potentially be habitable due to secondary 
factors, such as tidal heating \citep[e.g.,][]{bar09,hel13}, or could eventually become
habitable owing to the HZ's migration with increasing luminosity as the Sun ages
\citep[e.g.,][]{und03,jon05,gal17}.  Physical constraints on the likelihood of life on exoplanets
with a focus on stellar properties have been discussed by, e.g., \cite{lin18}.

Despite the fact that those example
cases are not considered classically habitable, they are of current interest 
to habitability even prompting future missions, such as the mission to
Europa, a moon of Jupiter, with the Europa Clipper\footnote{See {\tt https://www.jpl.nasa.gov/missions/europa-clipper}}
and to Saturn's largest moon, Titan, with
Dragonfly\footnote{See {\tt https://www.nasa.gov/press-release/nasas-dragonfly-will
-fly-around-titan-looking-for-origins-signs-of-life}}.
With this in mind, we are aware of the developing and ever-evolving habitability paradigms;
nonetheless, the scope of this study is solely devoted to the study of classical HZs.

Previous work by \cite{cun14,cun15}, focusing on HZs in stellar binary systems, 
demonstrated that binary star systems produce more intricate HZs than those of single 
star systems owing to orbital mechanics principles and orbital stability requirements.  Orbital 
mechanics allows stellar binary systems to possess two different types of orbits: the first being 
P-type which places the planet in an orbit external to and encircling both stars, whereas the 
second is S-type which places the planet in orbit around only one of the stars with the other 
acting as a perturber; see, e.g., \cite{dvo82} and a large body of subsequent work.

Though analysis of the possible HZs is necessary for our work, we focus primarily on the 
notion of dead zones (DZs) and their occurrence for various binary systems composed of main-sequence stars.  
The DZ is terminology given to represent parameter space where neither S-type nor 
P-type habitability is achievable; any system which falls within this region is deemed 
classically not habitable.  However, as said, just as candidate locations 
residing within classically defined HZs do not guarantee habitability, residency within 
defined DZs does not imply uninhabitability.  In the following, we present various sets
of model calculations pertaining to the existence of DZs.  Our paper is structured as follows:
Section~2 conveys the adopted theoretical approach, Section~3 presents our set of case studies,
Section~4 reports our discussion and conclusions, whereas Section~5 conveys an outlook also
considering future space missions.


\section{Theoretical approach}

In this study, we build on the previous work by \cite{cun14,cun15} and
\cite{wan19a}.  The adopted methodology includes both an analytical
and numerical approach allowing to determine S-type and P-type habitable
regions in stellar binary systems.  A joint constraint is applied including
the ensurance of orbital stability and a habitable region for a possible
system planet.  The first constraint consists in a gravitational criterion, whereas
the second constraint implies a radiative criterion; both of them in reference
to a possible system planet.

The radiative criterion allows to define a radiative habitable zone (RHZ)
as an intermediate step of the computational process.  In fact, the stellar
S-type and P-type RHZs in that approach are calculated through solving a
fourth-order polynomial.  Thereafter, considering the computational result
for the RHZs, the HZs can be identified.  Generally, they consist in subregions
of the RHZs, where according to the gravitational criterion orbital stability
for a possible system planet is warranted.  According to previous analyses,
e.g., \cite{cun14} and subsequent work, the HZ may encompass the entire domain
of the RHZ, parts of it, or it may not exist at all\footnote{As pointed out in
previous studies, combining constraints from both the RHZ and orbital stability limits can 
potentially truncate the resulting HZ for both S- and P-type orbits.  Following
\cite{cun14}, such truncations have been denoted as ST-type and PT-type, respectively.}.
Those outcomes are fully determined by the system's parameters.

There is also previous work by other authors on habitability in binary systems.
A recent up-to-date summary of research results is given by \cite{pil19} and
references therein.  Clearly, the prospect of habitability for planets and
moons in stellar binaries depends on a large set of conditions, including
(but not limited to) long-term orbital stability of the respective object,
the radiative environment as supplied by the stellar components, and the
likelihood of successful planet formation.

Another important aspect is the adopted planetary climate model.
In the following, we consider two different kinds of models previously
introduced in the literature.  One of those models allows to define the
general habitable zone (GHZ) previously studied by \cite{kop13,kop14}
(see below).  The other model allows to define the optimistic habitable zone
(OHZ) studied by the same authors; it is defined by the Recent Venus /
Early Mars limits (RVEM).  Fundamental work pertaining to both the GHZ
and OHZ has previously been pursued by \cite{kas93} and others; see,
e.g., \cite{kal17}, \cite{ram18}, and \cite{moo19} for further information
and applications to observed star-planet systems.

Regarding the GHZ, we adopt the approach pursued by \cite{kop14}.  It
includes parametric equations with climate model predicted inner and outer HZ
limit coefficients for calculating the GHZ for different planetary masses; i.e., 
Mars-mass (0.1~$M_{\oplus}$), Earth-mass ($1~M_{\oplus}$), and super-Earth-mass
(5~$M_{\oplus}$).  The convention also used here details the GHZ bounded by the
runaway greenhouse limit (inner limit) and the maximum greenhouse limit (outer limit).
Following \cite{kop14}, the GHZ can be calculated assuming
different planetary masses.

Hence, the investigation of the GHZ based on a Mars-mass, Earth-mass, and
super-Earth-mass planet assumption is henceforth denoted as GHZM, GHZE, and GHZSE,
respectively.  The OHZ, as defined above, however, does not depend on the planetary mass,
as pointed out by \cite{kop14}; thus, there is no analogous planetary mass distinction.
Lastly, we employ the work by \cite{man13} to convert mass and luminosity for the 
various stellar system components.  This approach has previously also been used
by \cite{wan19a,wan19b} and others.


\section{Case studies}

\subsection{Overview}

In order to explore how the associated HZs and DZs change in response to certain system
parameters, various case studies were chosen for investigation.  The first case study
details the scenario in which the stellar binary systems consist of identical stars (i.e.,
equal mass and spectral type).  The second case study applies an additional layer of complexity
such that the stellar primary mass is fixed while different values are assumed for the secondary mass
(and, by implication, luminosity).  The third case study compares HZs and the widths of the DZs
for selected equal-mass and nonequal-mass  systems while additionally comparing the outcomes for
Mars, Earth, and super-Earth planets. 


\subsection{Example~1: Equal mass and system plots --- spectral type representation}

The initial case study, intended as a tutorial example, examines how the HZs and DZs
change for stellar binaries with two identical stellar components.  Figure 1 illustrates
the width of the DZs as a function of stellar spectral type, binary semi-major axis 
$a_\mathrm{bin}$, and binary eccentricity $e_\mathrm{bin}$.  
In the four plots shown, both the primary and secondary stellar components 
share the same spectral type, given on the horizontal axis ranging from F0 to M0.  
In our investigation, we focus on main-sequence stars with luminosities moderately
above that of the Sun with an emphasis on low-luminosity stars, thus excluding
early-type stars.  This approach is justified by the shape of the initial mass function
that is significantly skewed toward low-mass stars \citep[e.g.,][]{kro01}.

The dark and light gray regions correspond to the GHZ and OHZ criterion, respectively. 
The results show that there is a DZ domain for each set of selected stellar parameters.
For relatively small values of $a_\mathrm{bin}$, P/PT-type HZs are observed, whereas
for relatively large values of $a_\mathrm{bin}$, S/ST-type HZs are identified, as
expected.  For fixed values of $e_\mathrm{bin}$, the extent of the DZs is largest for
high-luminosity stars and smallest for low-luminosity stars such as M-type stars (and
to a lesser degree for K-type stars).  Those systems of stars possess extended domains
of S-type habitability even for relatively small semi-major axes.

Another important aspect concerns the comparison of systems with different values of
$e_\mathrm{bin}$.  Here it is found that systems in circular orbits ($e_\mathrm{bin}
= 0.00$) possess the smallest DZ domains, both in reference to the GHZ and OHZ, whereas
highly eccentric systems ($e_\mathrm{bin} = 0.75$) possess the largest DZ domains.
This outcome is of course caused by the underlying assumptions that planets need
to remain in the HZs at all times without losing their prospects of habitability;
clearly, this condition is inapplicable to systems with planets of thick
atmospheres and/or other features in support of habitability as pointed out by,
e.g., \cite{wil02} and others.


\subsection{Example~2: Distance versus secondary mass examination}

The second case study examines how the HZs and DZs change as a function of 
secondary companion mass $M_{2}$ with the primary mass $M_{1}$ assumed as
fixed.  Figures 2, 3, and 4 convey the changing DZ 
width for primary masses of 0.60, 0.80, and 1.00 $M_{\odot}$, respectively, 
depicting each system at binary eccentricity values of 0.00, 0.25, 0.50, and 0.75.
As before, the gray regions correspond to the HZs, with dark and light gray corresponding 
to the GHZ and OHZ criteria, respectively, assuming an Earth-mass planet.  In addition,
we also report distinct corresponding values of luminosity $L_2$, in increments of
0.01~$L_\odot$ for $M_1 = 0.60~M_\odot$ and 0.10~$L_\odot$ for all other $M_1$ 
values, noting that $L_{2} = 1.00$ $L_\odot$ is coincidental with
$M_{2} = 1.00$ $M_\odot$.    

Figure~2 reports the case of $M_1 = 0.60~M_\odot$.    For the GHZ criterion,
the width of the DZ varies from 0.57~au to 0.74~au for $e_{\rm bin}$ = 0.00 and
from 4.39~au to 5.31~au for $e_{\rm bin}$ = 0.75 for the adopted range for secondary masses.
For the OHZ criterion, the DZ varies, correspondingly, from 0.39~au to 0.50~au for
$e_{\rm bin}$ = 0.00 and from 3.43~au to 4.15~au for $e_{\rm bin}$ = 0.75.
Moreover, Figure~4 conveys the case of $M_1 = 1.00~M_\odot$.  Here, for the GHZ criterion,
the width of the DZ varies from 2.02~au to 2.74~au for $e_{\rm bin}$ = 0.00 and
from 15.37~au to 19.32~au for $e_{\rm bin}$ = 0.75 for the adopted range for secondary masses.
For the OHZ criterion, the DZ varies, correspondingly, from 1.42~au to 1.90~au for
$e_{\rm bin}$ = 0.00 and from 12.04~au to 15.10~au for $e_{\rm bin}$ = 0.75.  These
results show that there is a notable increase in the DZ widths of both the GHZ and the OHZ
as the secondary mass is increased.  This behavior is due to the reduction of the
S/ST HZ as the stellar secondary components are of larger mass and, by implication,
larger luminosity.  This pattern is identified for any value of $e_{\rm bin}$.


\subsection{Example~3: Histograms with focus on the relevance of different planet masses}

The final case study examines the extent of the HZs and DZs for selected equal-mass and 
nonequal-mass systems.  Information is given in Figures 5 through 9 and in Tables 1, 2, and 3.
Figure 5 compares the different possible HZ regions for three equal-mass systems
($M_{1}$ = $M_{2}$ = $M$) where $M$ = 0.50, 0.75, and 1.00 $M_{\odot}$.
A particular aspect of this study is the distinguishedness regarding the GHZM, GHZE, and GHZSE,
which stem from the consideration of planets of different masses, which impact the
size and extent of the GHZ \citep{kop14}.  For each mass combination, we give two sets
of plots.  The left and right columns provide detailed plots depicting the P/PT-type and
S/ST-type HZ regions, respectively; plots in the left column are zoomed-in versions of
the P/PT-type HZ regions.  It is found that the DZs are modestly increased for
planets of smaller masses.

The dotted contours demonstrate the boundaries pertaining to the different 
HZ definitions.  For further illustration, the OHZ S/ST-type and P/PT-type HZ regions are
represented by the gray areas bounded by the dotted black line, whereas the GHZM S/ST-type
and P/PT-type HZ regions are represented by the gray areas bounded by the red dotted lines.
Furthermore, although the plots in the left column only show a small difference between the 
OHZ and the GHZs (i.e., GHZM, GHZE, GHZSE), if we were to zoom in further, an even smaller 
distinction between the different GHZs would be observable.  We also marked the differences
between the GHZs and OHZs; evidently, the DZs pertaining to the OHZs are well-pronounced
subregions of the DZs pertaining to the GHZs as the OHZs correspond to a wider range of
conditions potentially permitting exolife.

Figure 5 indicates that for equal-mass systems, the DZs are largest for systems
of largest stellar masses (i.e., also corresponding to highest luminosities).
Furthermore, the DZs are found to be increased for highly eccentric systems, i.e.,
high values of $e_{\rm bin}$.  This outcome is also consistent with the two previous
case studies (see Sect. 3.2 and 3.3).  Detailed numerical information is provided
in Table~1, where $e_{\rm bin}$ is varied in increments of 0.05.

Figure 6 compares the DZ widths for each system at specific binary eccentricities with 
different planetary mass assumptions where applicable.  For the GHZ criterion that includes 
the GHZM, GHZE, and GHZSE, the DZ extends from the bottom of each bar to different heights 
depending on the planetary mass assumption.  With a Mars-mass planet assumed, the corresponding 
DZ extends through the red portion exhibiting the largest DZ.  An Earth-mass planet assumption 
yields a DZ that extends through the blue portion but terminates where the red begins, resulting 
in the second largest DZ.  Further, a super-Earth-mass planet assumption would generate a DZ 
that extends through the green but terminates at the blue portion yielding a smaller DZ than the 
last two, and finally, the OHZ with any planetary mass assumption would yield the smallest DZ 
represented by the gray region which initiates slightly higher than the very bottom of the bar and 
terminates at the gray-green boundary. 

Figures 7 and 8 are presented in the same manner as Figs. 5 and 6, respectively; however, in this case, 
three nonequal-mass systems are investigated.  All systems chosen for investigation have the 
same secondary mass of $M_{2}$ = 0.50 $M_{\odot}$, but the primary masses vary as $M_{1}$ = 0.75, 
1.00, and 1.25 $M_{\odot}$.  Here the same pattern holds as found before:  the DZs are largest for systems
of largest stellar masses (i.e., also corresponding to highest luminosities).  Furthermore, the DZs are
found to be increased for highly eccentric systems, i.e., high values of $e_{\rm bin}$.  Detailed
numerical information is provided in Tables~2 and 3.  They also indicate that the family of GHZs
entail the largest DZ widths (in descending order: GHZM, GHZE, GHZSE).  Moreover, the DZ width
increases with increasing values of $e_{\rm bin}$, as expected; see also Tables 4 and 5 for
additional information.

Lastly, Fig. 9 compares six different stellar binary systems, half being equal-mass systems and 
the other half nonequal-mass systems, at $e_{\rm bin}$ = 0.00, 0.25, 0.50, and 0.75.  These 
plots deviate from the color coding of the previous plots in the sense that the DZ corresponding 
to the OHZ criterion is given in green whereas the DZs associated with the different GHZs are 
shown in varying shades of gray.  The darkest gray corresponds to a Mars-mass planet assumption 
and the lightest gray corresponds to a super-Earth-mass planet assumption, with the Earth-mass 
planet assumption lying in between in medium gray.   Strong tendencies are observed regarding
the DZs depending on the stellar masses.


\section{Discussion and conclusions}

The aim of this work encompassed case studies pertaining to classical habitability in stellar binary systems
with a particular emphasis on the description of DZs.  The theoretical approach is mostly based on the
previous contributions by \cite{wan19a}.  The adopted methodology is based on an intricate
approach allowing the determination of S/ST-type and P/PT-type habitable regions in stellar binary systems.  
Here we presented several sets of examples, reflective of systems with main-sequence stars of different
masses (and, by implication, luminosities), semi-major axes and eccentricities for the stellar components.
By examining those case studies, specific HZ and DZ trends became evident.  Our key findings include:

\smallskip\noindent
(1)  For systems of identical stars, the DZs are greatest in extent for relatively massive solar-type stars
(i.e., spectral type F) compared to less massive stars.  The smallest DZ extents are found for
pairs of M-type stars.

\smallskip\noindent
(2) Additionally, this trend is observed for each select binary eccentricity, with the DZ further increasing
with increasing binary eccentricity.  The smallest DZ occurs for stellar binary systems consisting of
low-mass stars in circular orbits (i.e., $e_{\rm bin}$ = 0.00).  Moreover, our investigations elucidate
the strong dependence of DZ width on S/ST-type habitability, with P/PT-type habitability exhibiting 
very little effect on the resulting DZ.

\smallskip\noindent
(3) Subsequent studies based on fixed primary stars but with a variable mass secondary companion
illustrate similar trends as pointed out in (1) and (2), with the DZ illustrating a positive correlation with binary 
eccentricity.  Numerical analyses show that the OHZ yields the smallest DZ.  This kind of pattern is also identified
for the GHZs in ascending order from super-Earth-mass planets to the largest DZ for Mars-mass planets. 

\smallskip\noindent
(4) Furthermore, the DZ increases with increasing primary mass and secondary mass, independently --- a behavior to 
be expected owing to the relationship between mass and luminosity (i.e., an increase in mass results in an increase 
in luminosity).  Our study also shows that DZ widths increase with increasing primary mass.  For the cases as studied,
the highest values for $M_{1}$ yield the largest DZ widths.

\smallskip\noindent
(5) Moreover, for the various kinds of GHZs as well as the OHZ, the width of the DZs notably increases as
a function of $e_{\rm bin}$.  This behavior is found for various sets of models, including those with fixed $M_1$
and variable values of $M_2$.  The increase in DZ width with increasing secondary mass is identified for each HZ criterion.

\smallskip\noindent
(6) We also compared the DZ widths for different HZ criteria for several stellar binary systems.  As expected, 
the OHZ yields a smaller DZ compared to each of the GHZs (GHZM, GHZE, and GHZSE) for all primary / secondary 
mass combinations over the entire range of eccentricities.  Previously, additional information pertaining to
intermediate planetary masses has been given by \cite{wan19b}.

\smallskip

We limited our investigation to main-sequence stars with masses and luminosities between
spectral types F and M.  The justification for this restriction lies in the well-known
initial mass function  (IMF), illustrating that lower mass, and therefore lower luminosity,
stars occur in higher frequency \citep[][ and subsequent work]{kro01, kro02,cha03}.
Additionally, low-mass / low-luminosity stars appear to possess some preferential
features to support environments favorable for life, possibly including advanced life,
and may also more likely permit exolife detectability \citep[e.g.,][]{hel14,cun16,arn19}.


\section{Outlook}

Future planetary search mission such as the
{\it Transiting Exoplanet Survey Satellite}
(TESS)\footnote{See {\tt https://www.nasa.gov/tess-transiting-exoplanet-survey-satellite}}, the
{\it James Webb Space Telescope} (JWST)\footnote{See {\tt https://www.jwst.nasa.gov}}, the
{\it CHaracterising ExOPlanets Satellite} (CHEOPS)\footnote{See {\tt https://sci.esa.int/web/cheops}},
and the space telescope {\it PLAnetary Transits and Oscillations of stars}
(PLATO)\footnote{See {\tt https://sci.esa.int/web/plato}},
are expected to shed further light on the existence and characteristics of extrasolar planets,
including planets in DZs of stellar binaries.  Generally, DZs still allow the existence of
planets in those zones noting that they are expected to contain subzones permitting planetary
orbital stability.  The latter are expected to occur in close proximity to stars for S-type
orbits and beyond considerable distances from the center of mass for P-type orbits; see, e.g.,
\cite{qua18} for detailed planetary stability studies.

TESS will search for exoplanets that periodically block part of the light from their host stars,
events called transits.  TESS is expected to survey 200,000 of the brightest stars in the
solar neighborhood.  JWST is capable to answer numerous questions pertaining to planetary
science, including determining the building blocks of planets, the constituents of
circumstellar disks, and various aspects of planetary evolution relevant to astrobiology
and astrochemistry.  The main goal of CHEOPS is the measurement of planetary radii, allowing
to describe the planets' densities and approximate compositions.  PLATO, on the other hand
(if launched), will focus on planetary transits across up to one million stars to discover and
characterize rocky exoplanets around Sun-like stars, subgiants, and K and M dwarfs.

It is noteworthy, however, that for planets situated in DZs the radiative (climatological)
criterion as defined for classical habitability (see Section 2) will not be met; hence,
the reason for the DZ.  Therefore, possible life forms (if existing) might be based on
non-terrestrial biochemistry \citep[e.g.,][]{bai04}; note that the latter may include
``ammonochemistry'' and ``silicon biochemistry'' with the latter potentially also present
in the Solar System planetary-sized objects beyond Mars.  Ultimately, additional observational
campaigns are desired.  Those should also encompass the search for biosignatures, including
molecules and molecular disequilibria that are apparently irreproducible through non-biological
processes; see, e.g., \cite{sea16}, \cite{mea18}, \cite{ols18}, \cite{sch18}, and \cite{arn19}
for background information.


\section*{Acknowledgments}

This work has been supported by the Department of Physics,
University of Texas at Arlington and the U.S. Department of 
Education through the Graduate Assistance in Areas of National 
Need (GAANN) program (S.~Y.~M.).  Additionally, we wish
to draw the reader's attention to the online tool {\tt BinHab 2.0},
created by one of us (Zh.~W.) and hosted at The University of Texas at Arlington.
It allows the calculation of habitable regions in binary systems
based on the developed method.





%
%
\begin{table}
\caption{DZ Widths for $M_{1}$ = $M_{2}$ = 1.00 $M_{\odot}$}
\centerline{\begin{tabular}{c c c c c} \hline
\noalign{\smallskip}
$e\rm_{bin}$ & GHZM & GHZE & GHZSE & OHZ \\
...          & (au) & (au) & (au)  & (au) \\
\noalign{\smallskip}
\hline
\hline
\noalign{\smallskip}
$0.00$		& 2.94		&	2.73			&	2.60		& 1.90	 \\			    
$0.05$		& 3.25		&	3.03			&	2.89		& 2.14   \\				
$0.10$		& 3.58		&	3.34			&	3.19		& 2.40   \\
$0.15$		& 3.93		&	3.67			&	3.52		& 2.69   \\
$0.20$		& 4.32		&	4.05			&	3.88		& 2.98   \\
$0.25$		& 4.76		&	4.46			&  4.28		& 3.32	 \\
$0.30$		& 5.26		&	4.94			&	4.74		& 3.71   \\
$0.35$		& 5.84		&	5.49			&	5.27		& 4.14   \\
$0.40$		& 6.51		&	6.12			&	5.89		& 4.65   \\
$0.45$		& 7.32		&	6.88			&	6.62		& 5.26   \\
$0.50$		& 8.30		&	7.81			&  7.52		& 5.99	 \\
$0.55$		& 9.52		&	8.97			&	8.63		& 6.91   \\
$0.60$		& 11.08		&	10.45		&	10.07	& 8.09   \\
$0.65$		& 13.17		&	12.42		&	11.98	& 9.65   \\
$0.70$		& 16.10		&	15.20		&	14.65	& 11.84   \\
$0.75$		& 20.51		&	19.37		&  18.68	& 15.13	   \\
$0.80$		& 27.84		&	26.30		&	25.36	& 20.61   \\
\hline
\end{tabular}
}
\end{table}

%
%
\begin{table}
\caption{DZ Widths for $M_{1}$ = 1.00 $M_{\odot}$, $M_{2}$ = 0.75 $M_{\odot}$}
\centerline{\begin{tabular}{c c c c c} \hline
\noalign{\smallskip}
$e\rm_{bin}$ & GHZM & GHZE & GHZSE & OHZ \\
...          & (au) & (au) & (au)  & (au) \\
\noalign{\smallskip}
\hline
\hline
\noalign{\smallskip}
$0.00$		& 2.66		&	2.47			&	2.36		& 1.74	 \\				
$0.05$		& 2.95		&	2.74			&	2.62		& 1.96   \\				
$0.10$		& 3.23		&	3.02			&	2.89		& 2.20   \\
$0.15$		& 3.56		&	3.33			&	3.19		& 2.44   \\
$0.20$		& 3.91		&	3.66			&	3.51		& 2.72   \\
$0.25$		& 4.31		&	4.05			&  3.88		& 3.03	 \\
$0.30$		& 4.76		&	4.48			&	4.30		& 3.38   \\
$0.35$		& 5.30		&	4.98			&	4.78		& 3.78	 \\
$0.40$		& 5.91		&	5.56			&	5.35		& 4.24   \\
$0.45$		& 6.65		&	6.26			&	6.02		& 4.80   \\
$0.50$		& 7.55		&	7.11			&  6.85		& 5.48	 \\
$0.55$		& 8.68		&	8.18			&	7.88		& 6.32   \\
$0.60$		& 10.12		&	9.55			&	9.20		& 7.40   \\
$0.65$		& 12.05		&	11.37		&	10.96	& 8.85    \\
$0.70$		& 14.76		&	13.93		&	13.43	& 10.87	   \\
$0.75$		& 18.85		&	17.80		&  17.16	& 13.93    \\
$0.80$		& 25.64		&	24.22		&	23.37	& 19.00    \\
\hline
\end{tabular}
}
\end{table}

%
%
\begin{table}
\caption{DZ Widths for $M_{1}$ = 1.00 $M_{\odot}$, $M_{2}$ = 0.50 $M_{\odot}$}
\centerline{\begin{tabular}{c c c c c} \hline
\noalign{\smallskip}
$e\rm_{bin}$ & GHZM & GHZE & GHZSE & OHZ \\
...          & (au) & (au) & (au)  & (au) \\
\noalign{\smallskip}
\hline
\hline
\noalign{\smallskip}
$0.00$		& 2.35		&	2.18			&	2.08			& 1.54   \\				
$0.05$		& 2.60		&	2.43			&	2.32			& 1.75   \\				
$0.10$		& 2.86		&	2.68			&	2.56			& 1.96   \\
$0.15$		& 3.15		&	2.95			&	2.83			& 2.19   \\
$0.20$		& 3.48		&	3.26			&	3.13			& 2.44   \\
$0.25$		& 3.84		&	3.60			&   3.46			& 2.71	 \\
$0.30$		& 4.25		&	3.99			&	3.84			& 3.03   \\
$0.35$		& 4.73		&	4.45			&	4.28			& 3.39   \\
$0.40$		& 5.28		&	4.98			&	4.78			& 3.81   \\
$0.45$		& 5.96		&	5.61			&	5.40			& 4.31   \\
$0.50$		& 6.78		&	6.38			&  6.15			& 4.93	 \\
$0.55$		& 7.80		&	7.36			&	7.08			& 5.70   \\
$0.60$		& 9.12		&	8.60			&	8.29			& 6.69   \\
$0.65$		& 10.88		&	10.27		&	9.90			& 8.00   \\
$0.70$		& 13.36		&	12.61		&	12.16		& 9.85   \\
$0.75$		& 17.09		&	16.14		&  15.57		& 12.65	  \\
$0.80$		& 23.29		&	22.00		&	21.23		& 17.28   \\
\hline
\end{tabular}
}
\end{table}

%
%
\begin{table}
\caption{DZ Widths for different $M_{1}$ with $M_{2}$ = 0.60 $M_{\odot}$}
\centerline{\begin{tabular}{c c c c c} \hline
\noalign{\smallskip}
$M_{1}$       & GHZE & GHZE & OHZ  & OHZ  \\
($M_{\odot}$) & (au) & (au) & (au) & (au) \\
\noalign{\smallskip}
\hline
\noalign{\smallskip}
...           & $e\rm_{bin}$ = 0.00 & $e\rm_{bin}$ = 0.75 & $e\rm_{bin}$ = 0.00 & $e\rm_{bin}$ = 0.75 \\
\noalign{\smallskip}
\hline
\hline
\noalign{\smallskip}
$0.60$		& 0.74		&	5.31			&	0.50		&  4.15  \\				
$0.80$		& 1.85		&	13.06		&	1.33		&  10.23  \\				
$1.00$		& 2.33		&	16.75		&	1.67		&  13.12  \\
\hline
\end{tabular}
}
\end{table}

%
%
\begin{table}
\caption{DZ Widths for different $M_{1}$ with $M_{2}$ = 0.40 $M_{\odot}$}
\centerline{\begin{tabular}{c c c c c} \hline
\noalign{\smallskip}
$M_{1}$       & GHZE & GHZE & OHZ  & OHZ  \\
($M_{\odot}$) & (au) & (au) & (au) & (au) \\
\noalign{\smallskip}
\hline
\noalign{\smallskip}
...           & $e\rm_{bin}$ = 0.00 & $e\rm_{bin}$ = 0.75 & $e\rm_{bin}$ = 0.00 & $e\rm_{bin}$ = 0.75 \\
\noalign{\smallskip}
\hline
\hline
\noalign{\smallskip}
$0.60$		& 0.57		&	4.39			&	0.39		&  3.43    \\				
$0.80$		& 1.60		&	11.87		&	1.13		&  9.30  \\				
$1.00$		& 2.02		&	15.37		&	1.42		&  12.04  \\
\hline
\end{tabular}
}
\end{table}

%
%
\begin{figure*}
\centering
\begin{tabular} {c}
\includegraphics[width=0.5\linewidth]{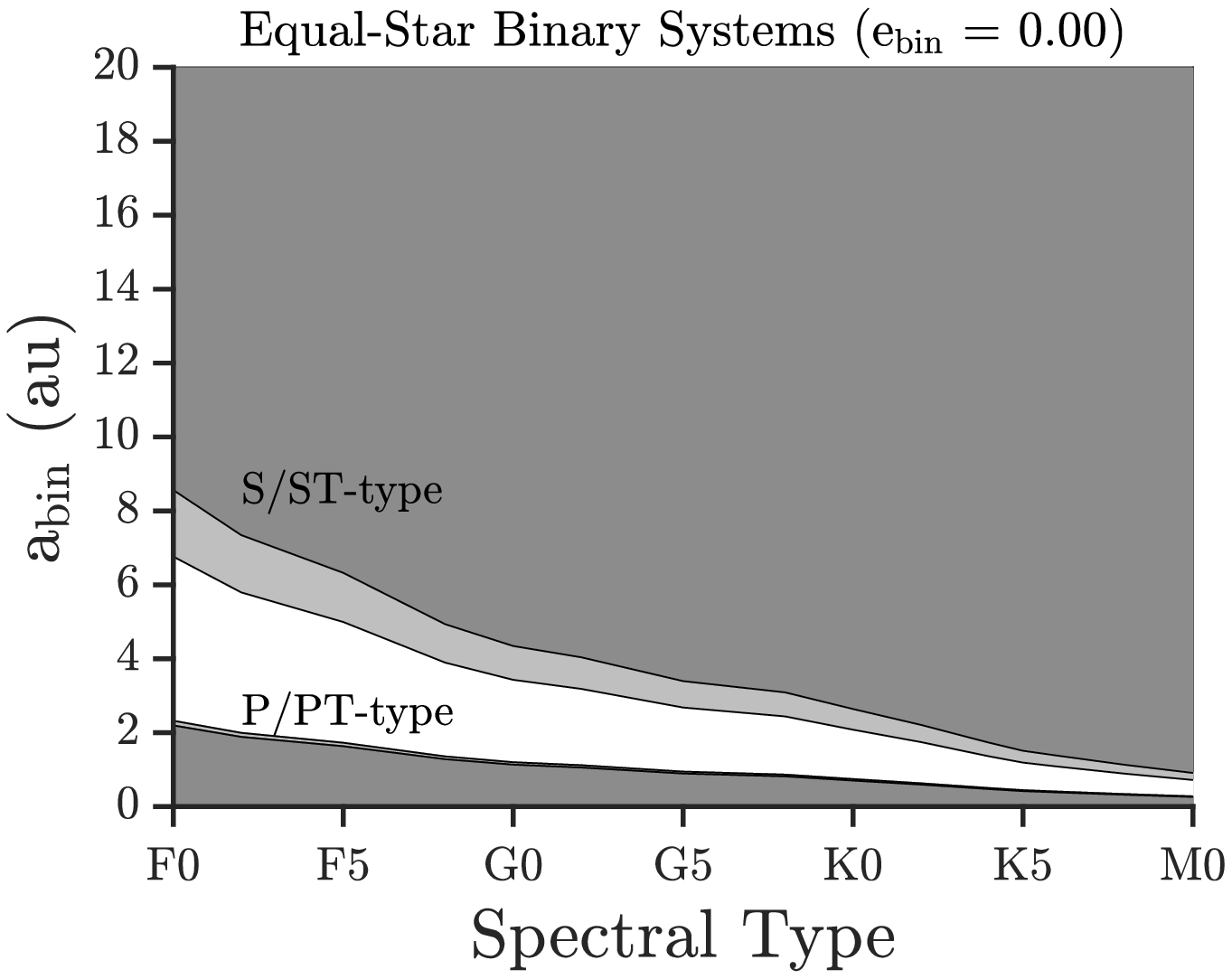}
\hspace{0.5cm}
\includegraphics[width=0.5\linewidth]{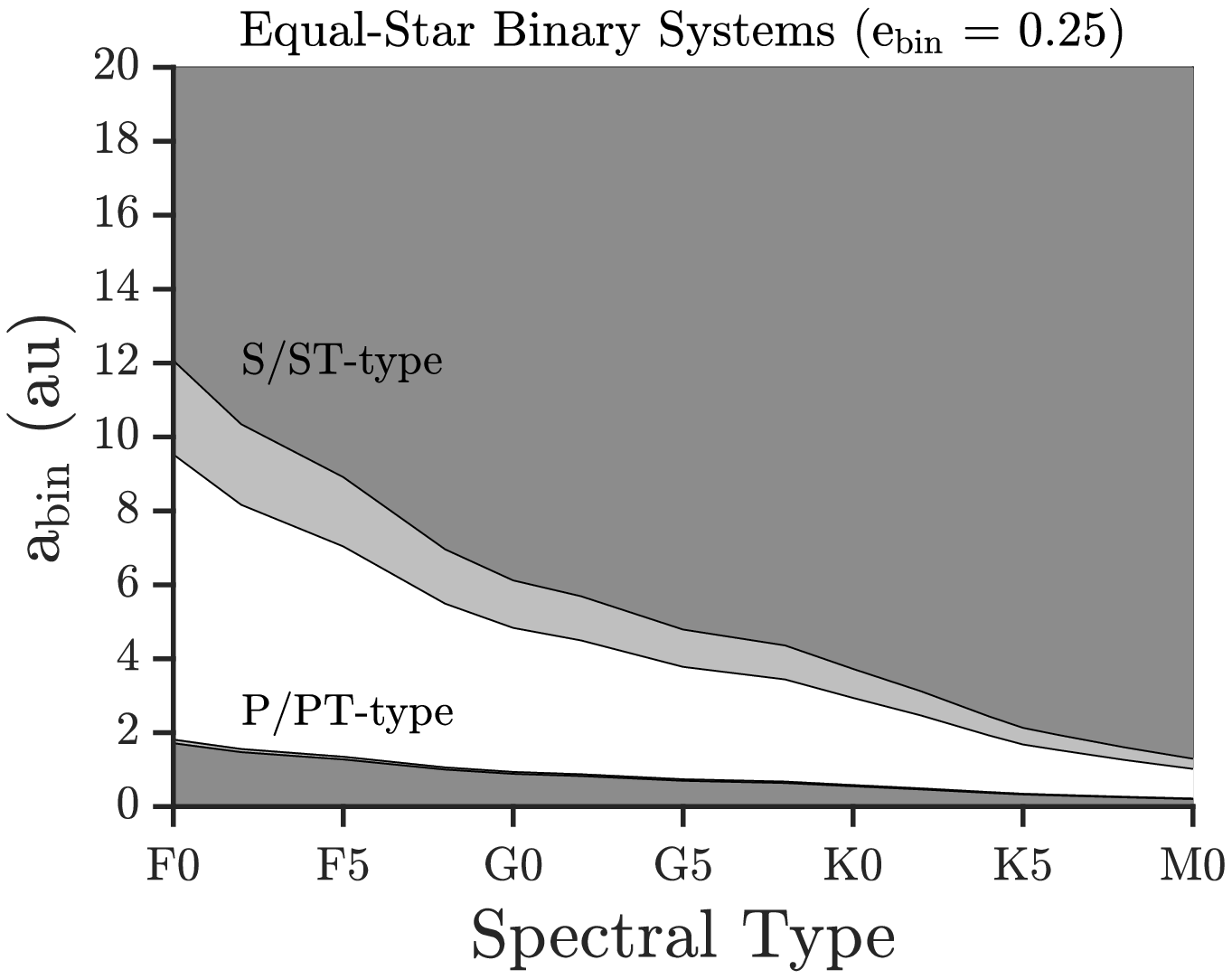}
\\
\\
\includegraphics[width=0.5\linewidth]{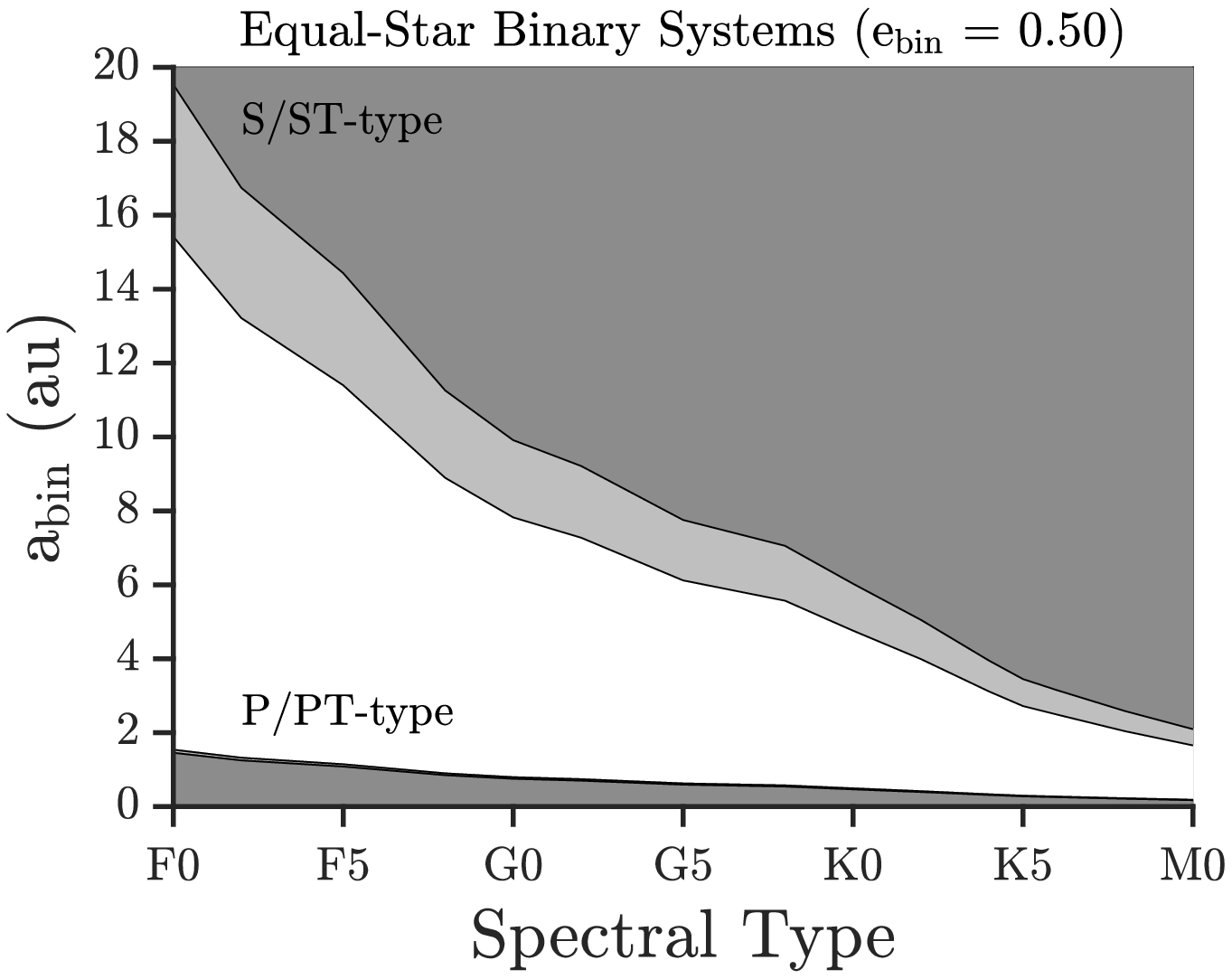}
\hspace{0.5cm}
\includegraphics[width=0.5\linewidth]{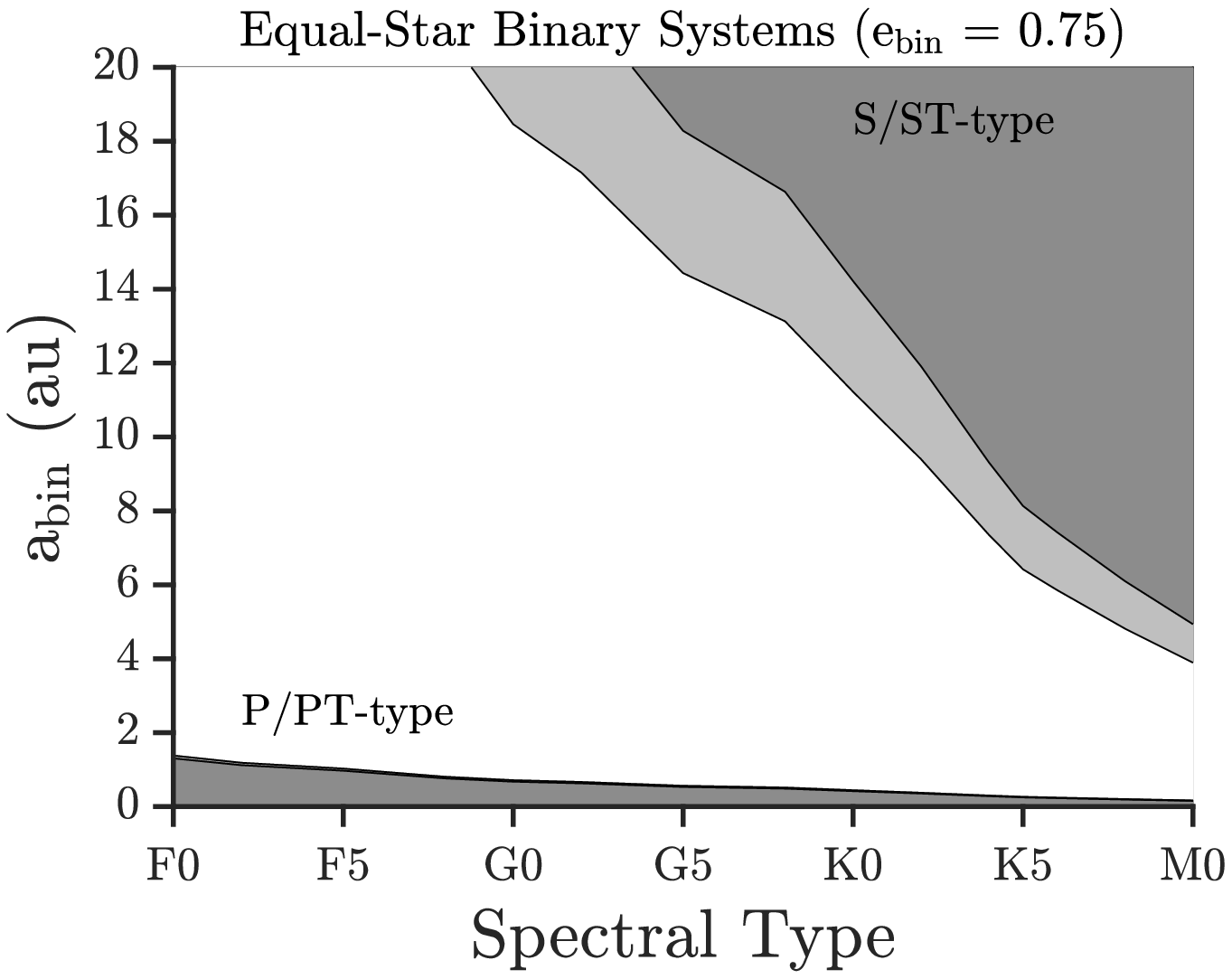} \\
\end{tabular}
\vspace{1cm}
\caption{
Case study 1: Calculated HZs and DZs for equal-mass (and identical spectral type) stellar binary systems 
($M_{1} =$ $M_{2} =$ $M$) for different masses, $M$, corresponding to spectral types ranging from 
F0 to M0 at selected binary eccentricities ($e_{\rm bin}$ = 0.00, 0.25, 0.50, and 0.75).  The gray regions 
indicate the different possible HZs (i.e., S/ST-type and P/PT-type) with the white regions in between illustrating 
the DZs.  The dark and light gray regions correspond to the adopted definitions of GHZ and OHZ, respectively, 
assuming an Earth-mass planet. 
}
\end{figure*}

%
%
\begin{figure*}
\centering
\begin{tabular} {c}
\includegraphics[width=0.5\linewidth]{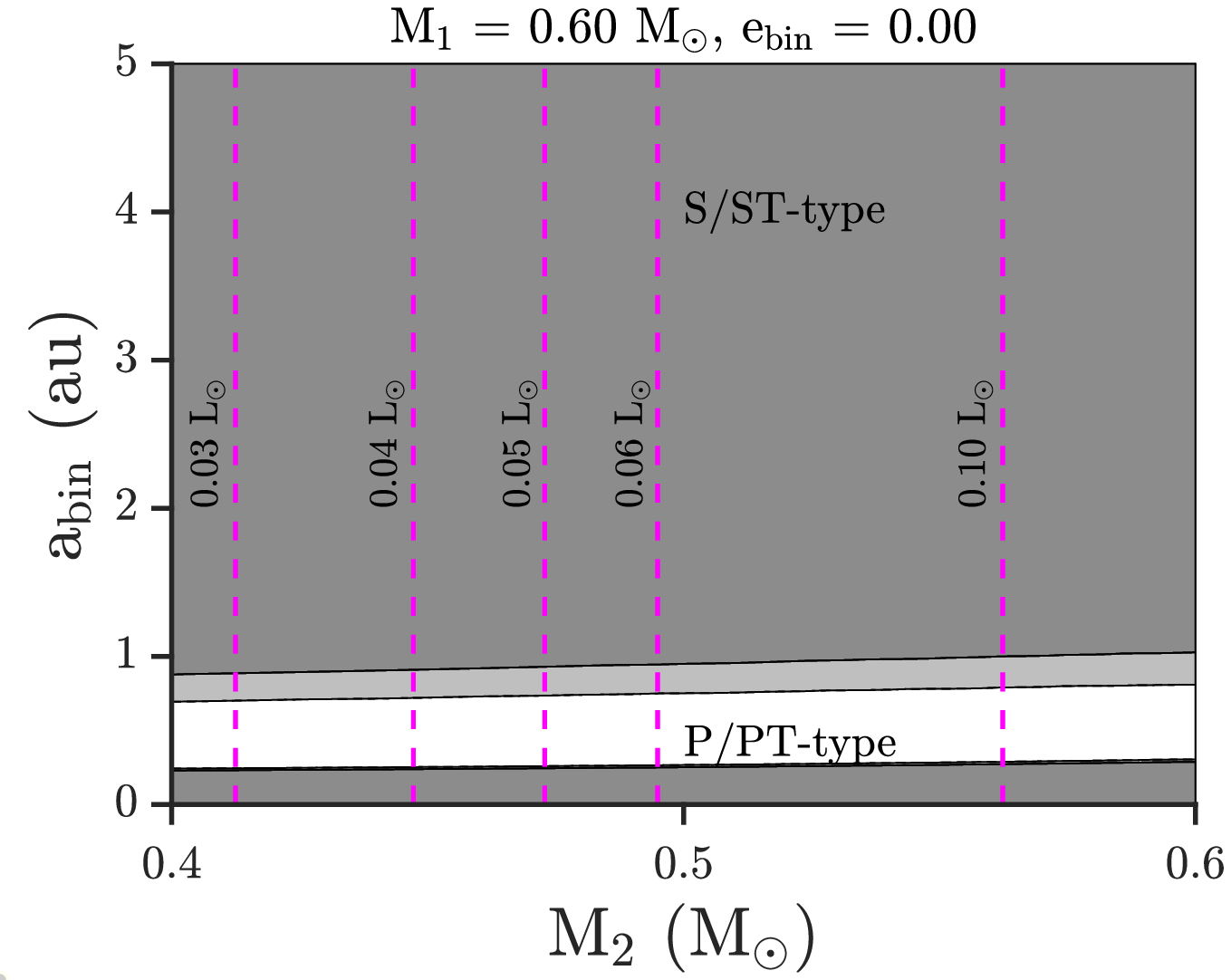}
\hspace{0.5cm}
\includegraphics[width=0.5\linewidth]{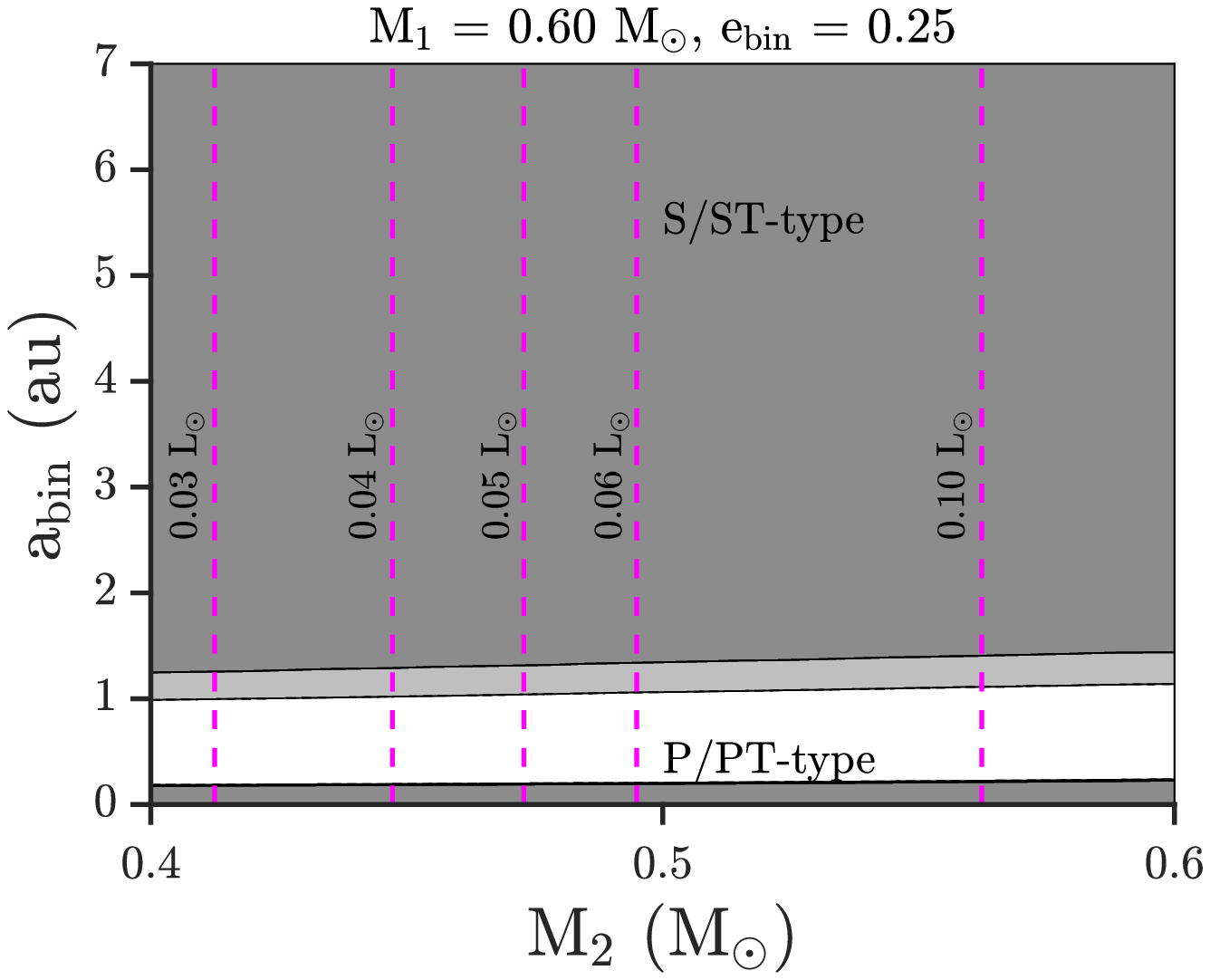}
\\
\\
\includegraphics[width=0.5\linewidth]{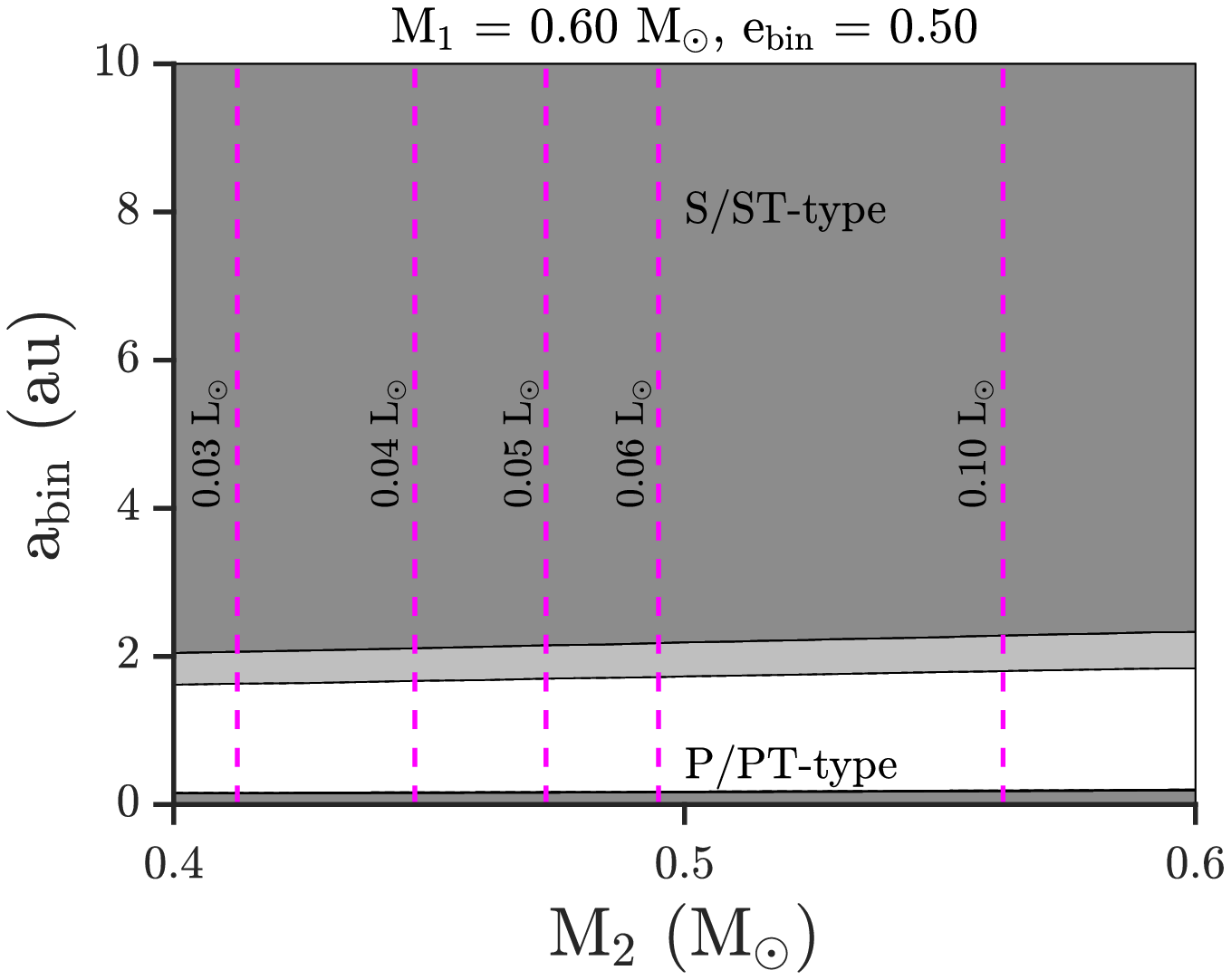}
\hspace{0.5cm}
\includegraphics[width=0.5\linewidth]{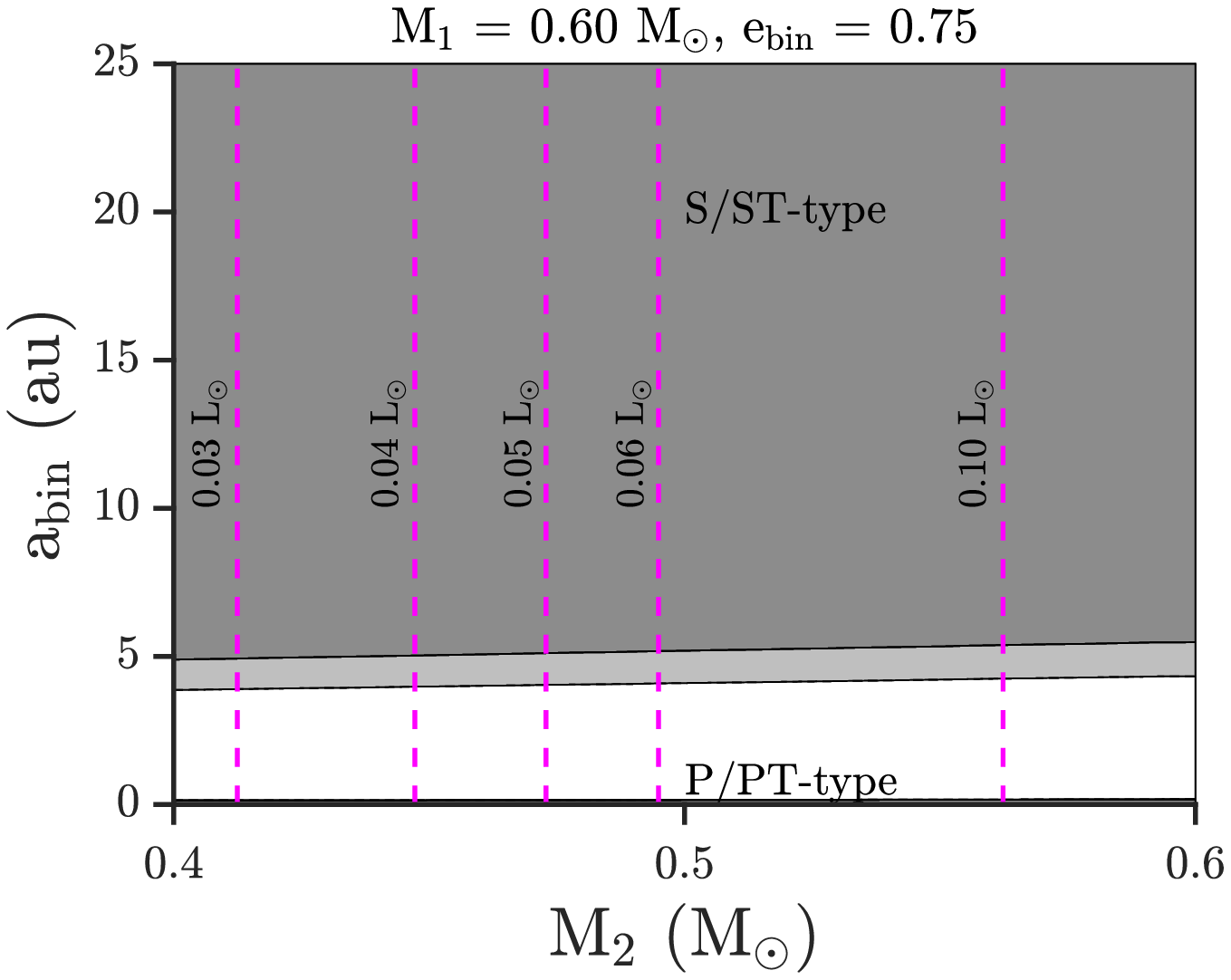} \\
\end{tabular}
\vspace{1cm}
\caption{
Case study 2: Calculated HZs and DZs for a stellar binary system consisting of a primary 
($M_{1} = 0.60$ $M_\mathrm{\odot}$) with a variable mass secondary companion ($M_{2}$) at 
selected binary eccentricities ($e_{\rm bin}$ = 0.00, 0.25, 0.50, and 0.75).  The gray regions indicate 
the different possible HZs (i.e., S/ST-type and P/PT-type) with the white regions in between illustrating the 
DZs.  The dark and light gray regions correspond to the adopted definitions of GHZ and OHZ, respectively, 
assuming an Earth-mass planet.  The vertical dotted lines (magenta) convey lines of 
secondary companion luminosity, $L_{2}$.
}
\end{figure*}

%
%
\begin{figure*}
\centering
\begin{tabular} {c}
\includegraphics[width=0.5\linewidth]{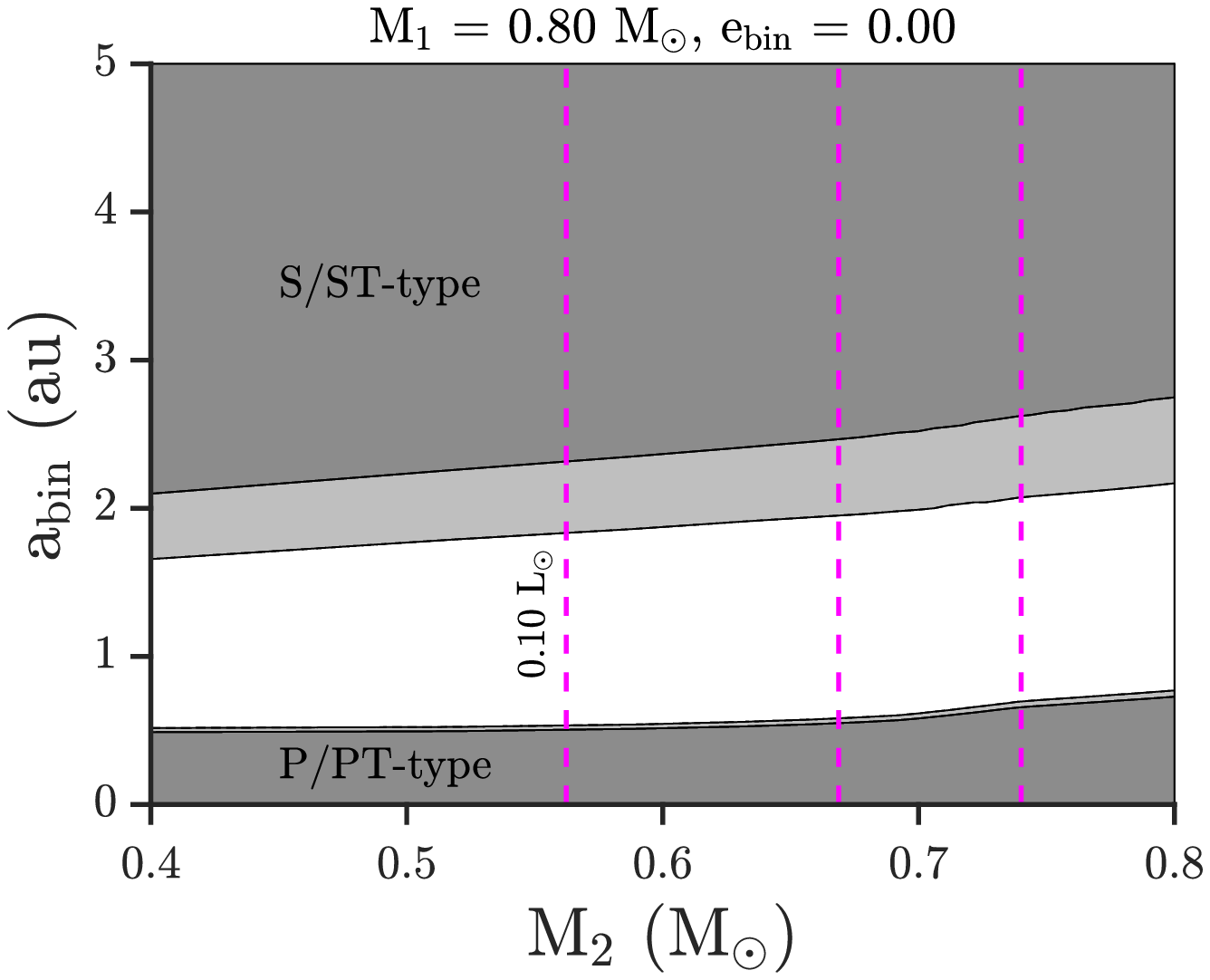}
\hspace{0.5cm}
\includegraphics[width=0.5\linewidth]{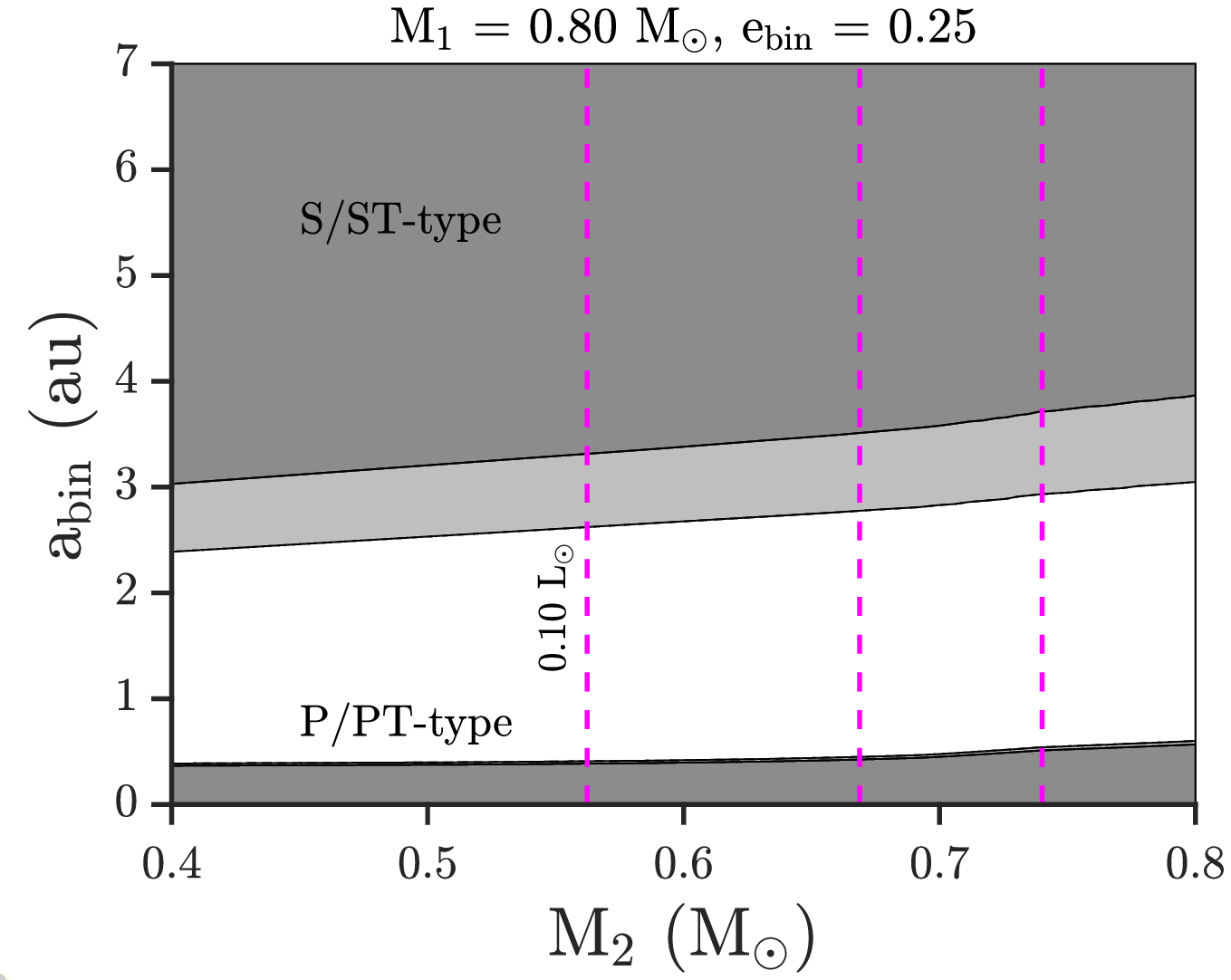}
\\
\\
\includegraphics[width=0.5\linewidth]{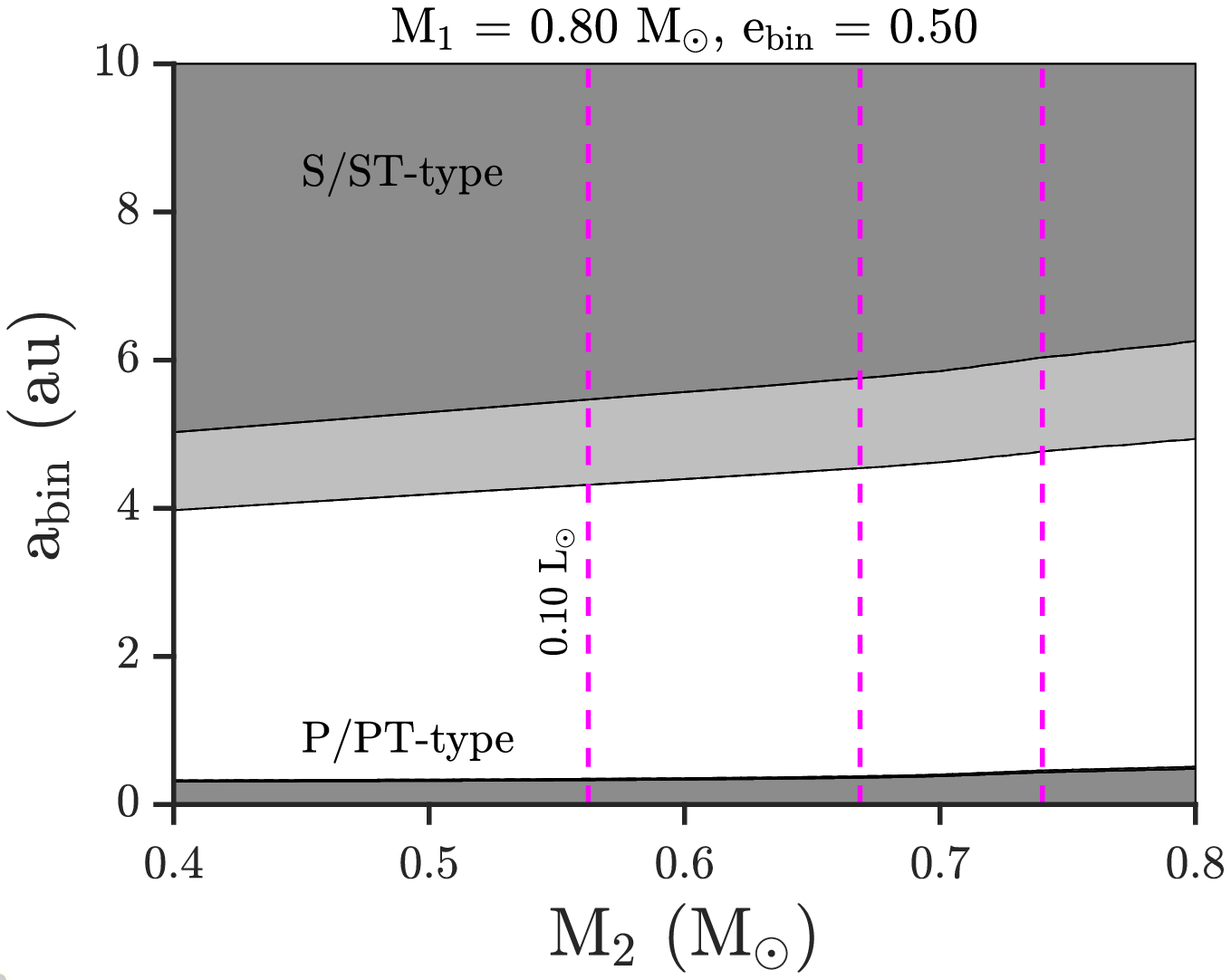}
\hspace{0.5cm}
\includegraphics[width=0.5\linewidth]{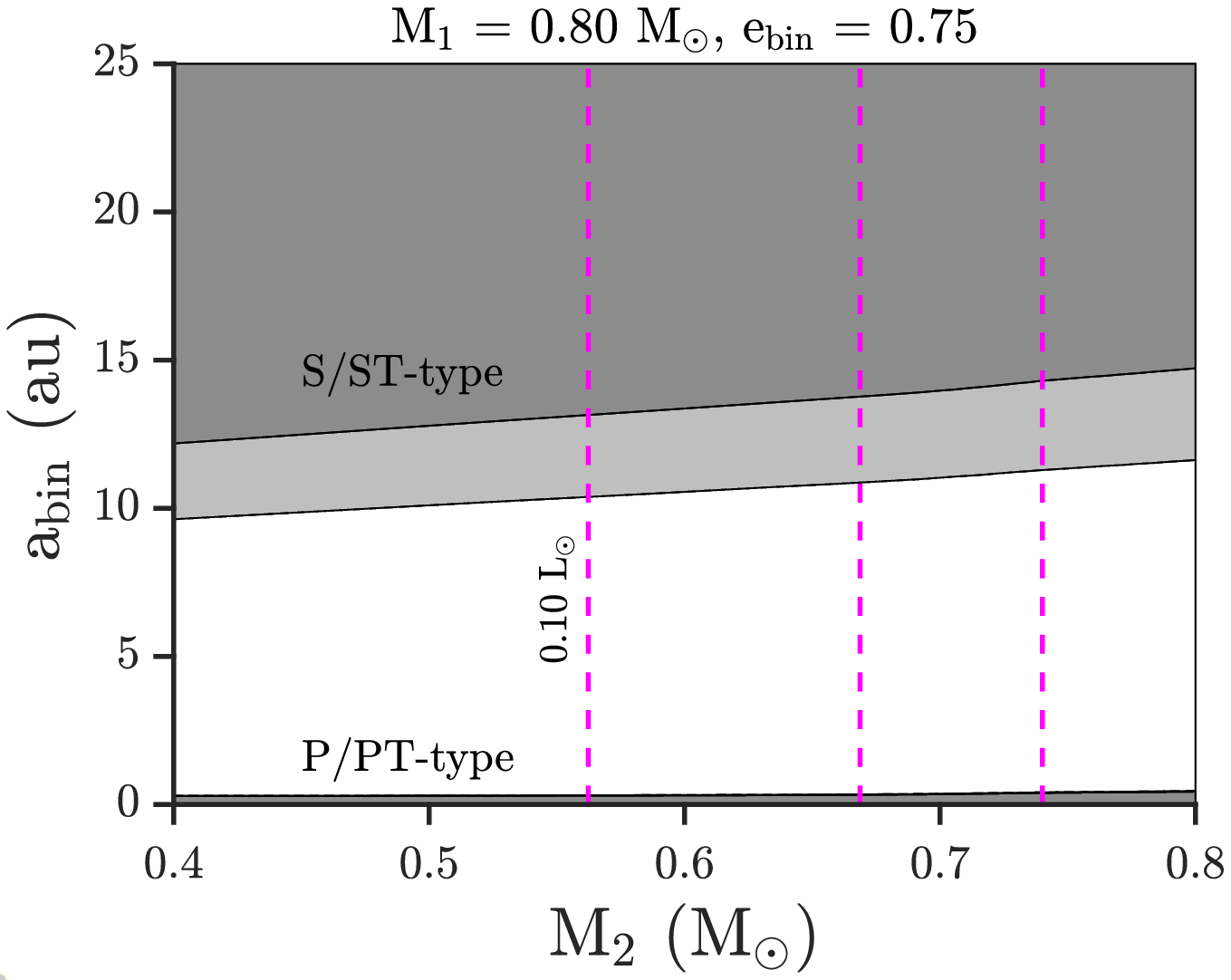} \\
\end{tabular}
\vspace{1cm}
\caption{
Case study 2: Same as Fig.~2, but for a primary of $M_{1} = 0.80$ $M_\mathrm{\odot}$ and 
secondary companion luminosity, $L_{2}$, in increments of 0.10~$L_{\mathrm{\odot}}$, with the
luminosity increasing from left to right.
}
\end{figure*}

%
%
\begin{figure*}
\centering
\begin{tabular} {c}
\includegraphics[width=0.5\linewidth]{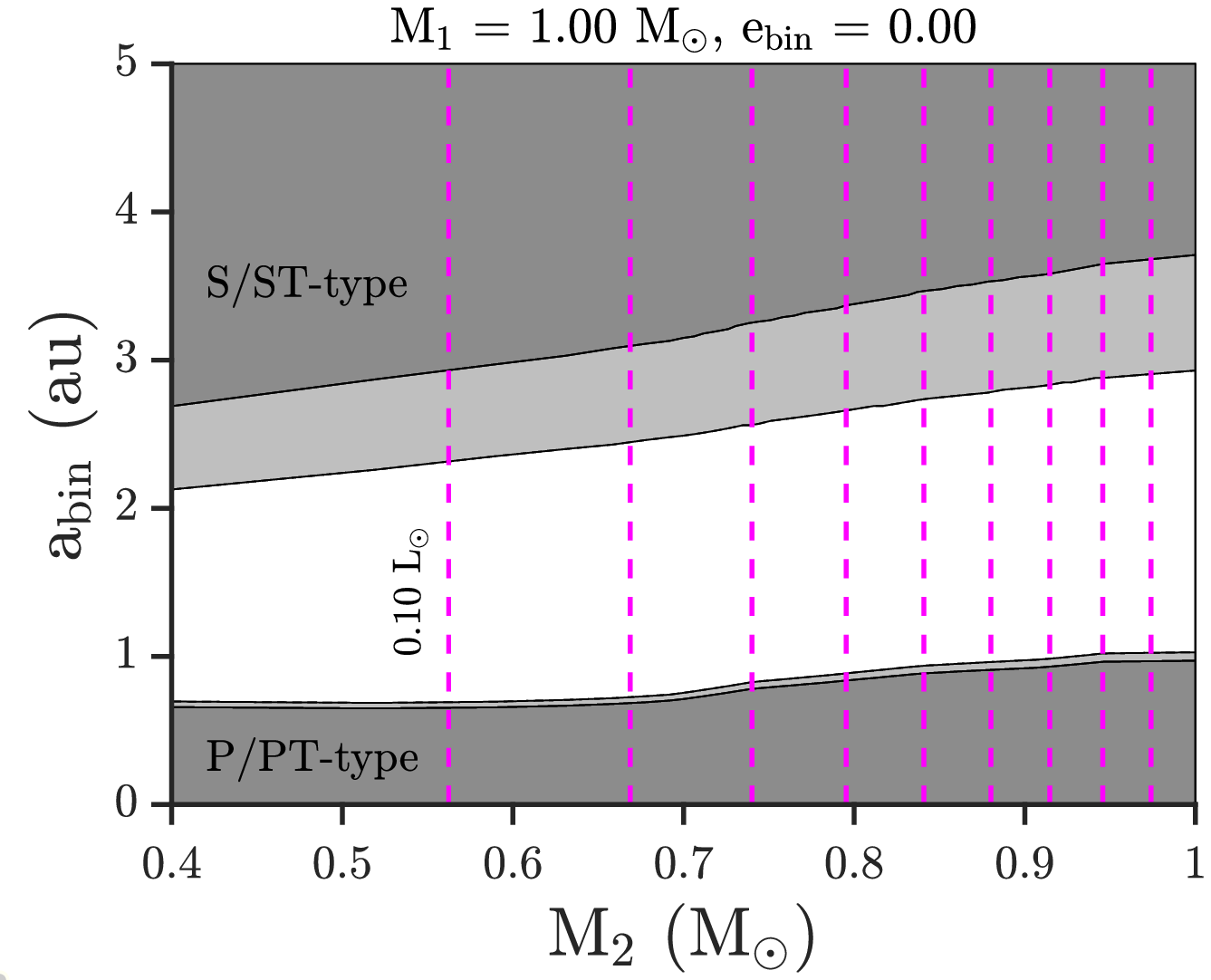}
\hspace{0.5cm}
\includegraphics[width=0.5\linewidth]{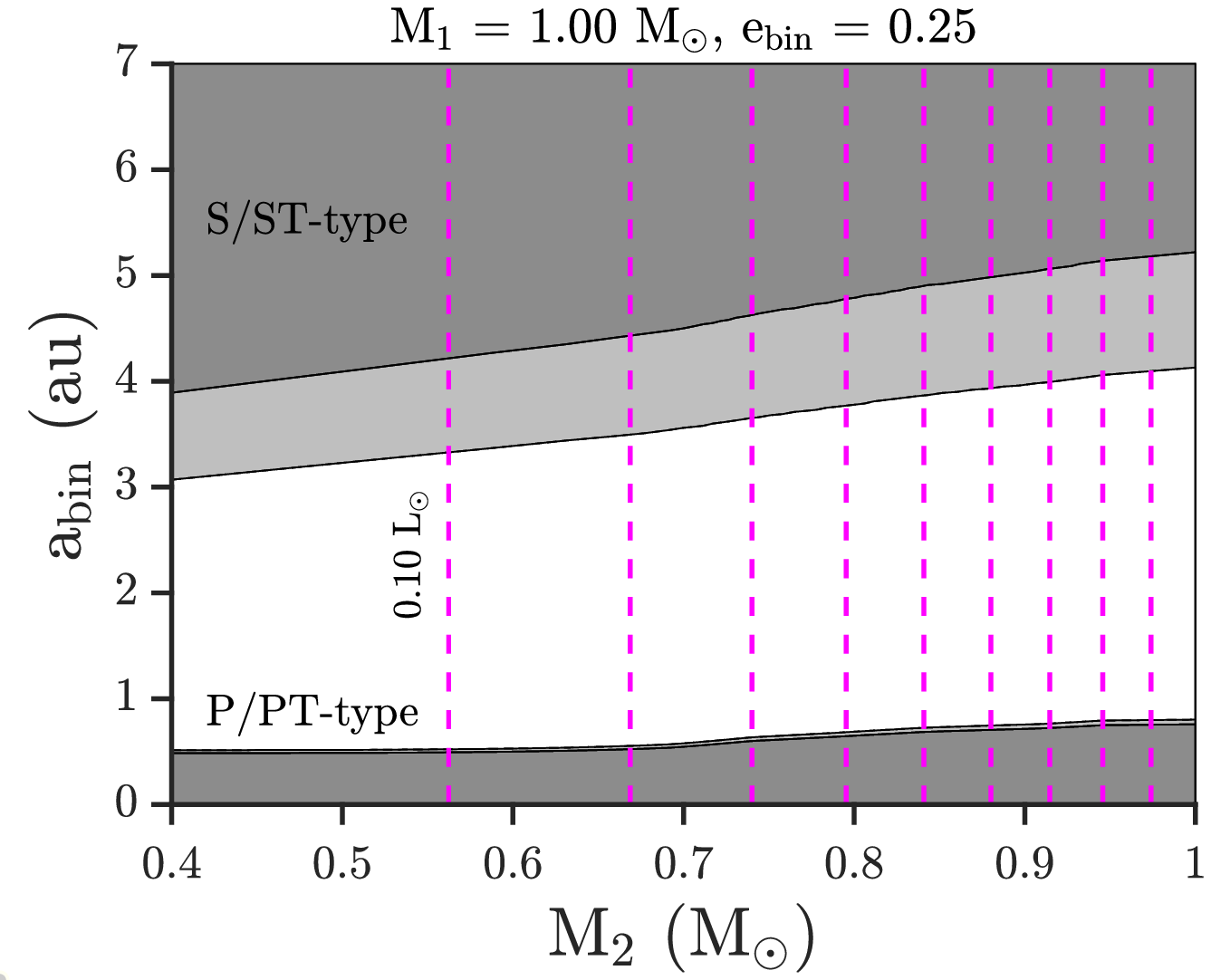}
\\
\\
\includegraphics[width=0.5\linewidth]{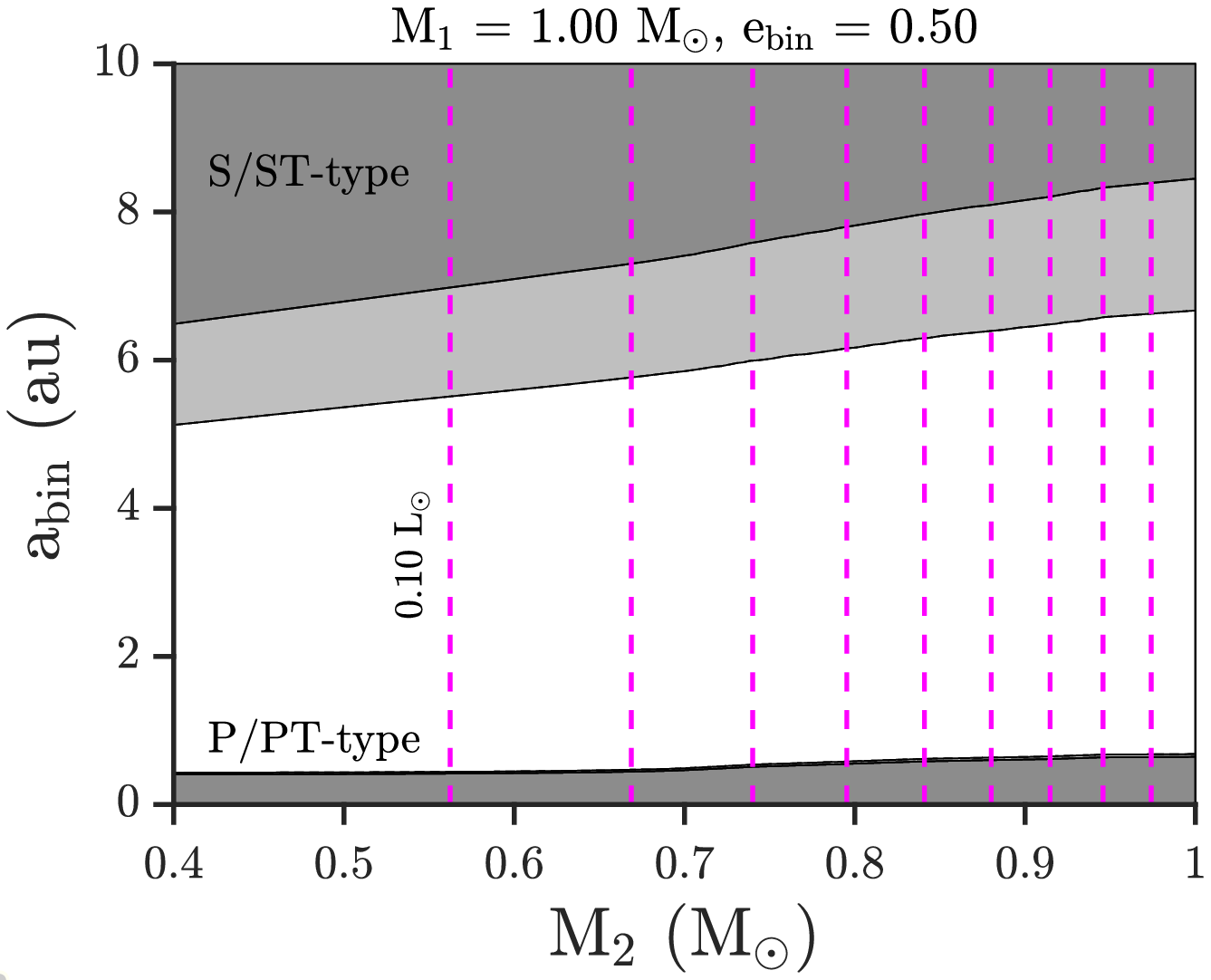}
\hspace{0.5cm}
\includegraphics[width=0.5\linewidth]{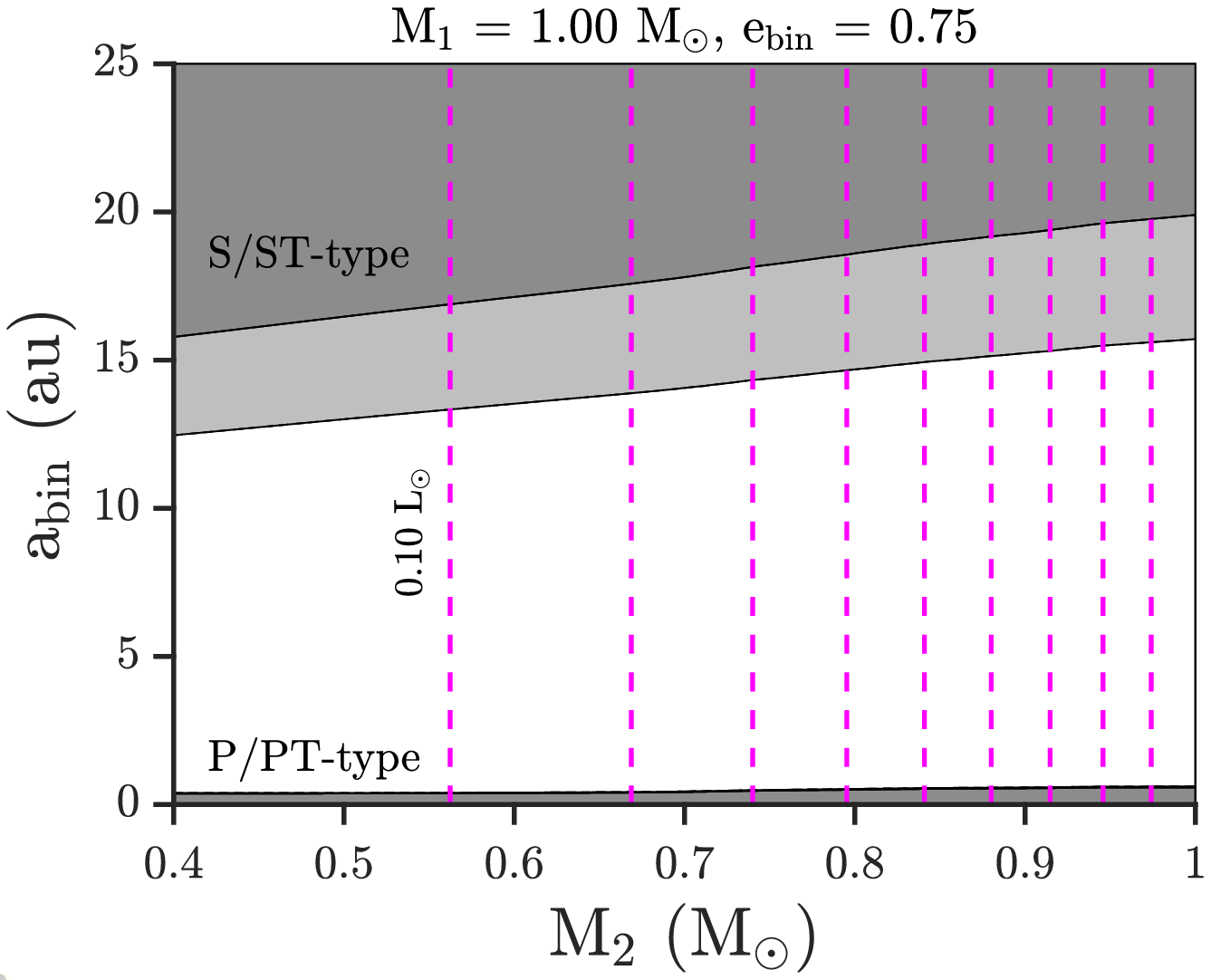} \\
\end{tabular}
\vspace{1cm}
\caption{
Case study 2: Same as Fig.~3, but for a primary of $M_{1} = 1.00$ $M_\mathrm{\odot}$.
}
\end{figure*}

%
%
\begin{figure*}
\centering
\begin{tabular} {c}
\includegraphics[width=0.45\linewidth]{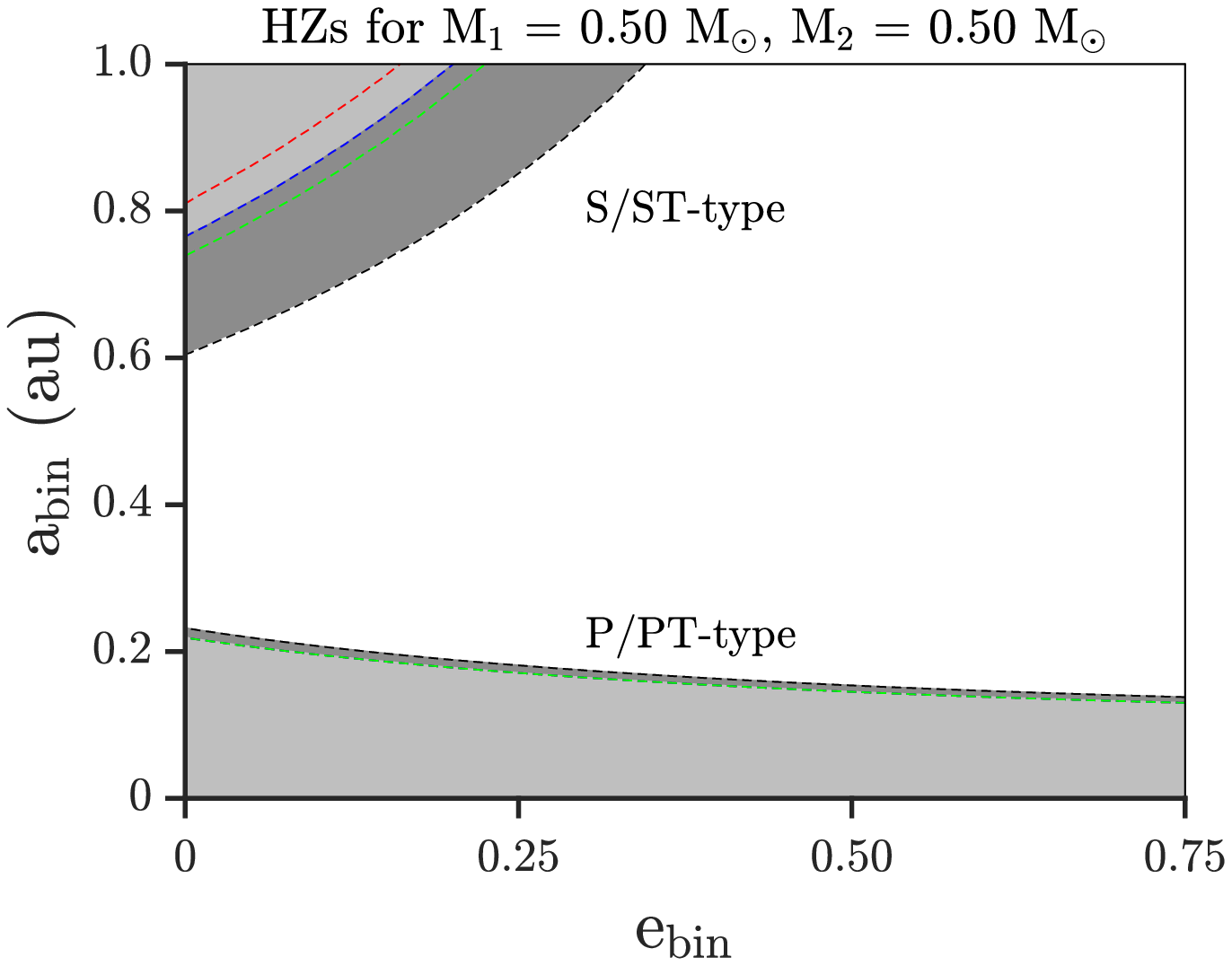} 
\hspace{0.5cm}
\includegraphics[width=0.45\linewidth]{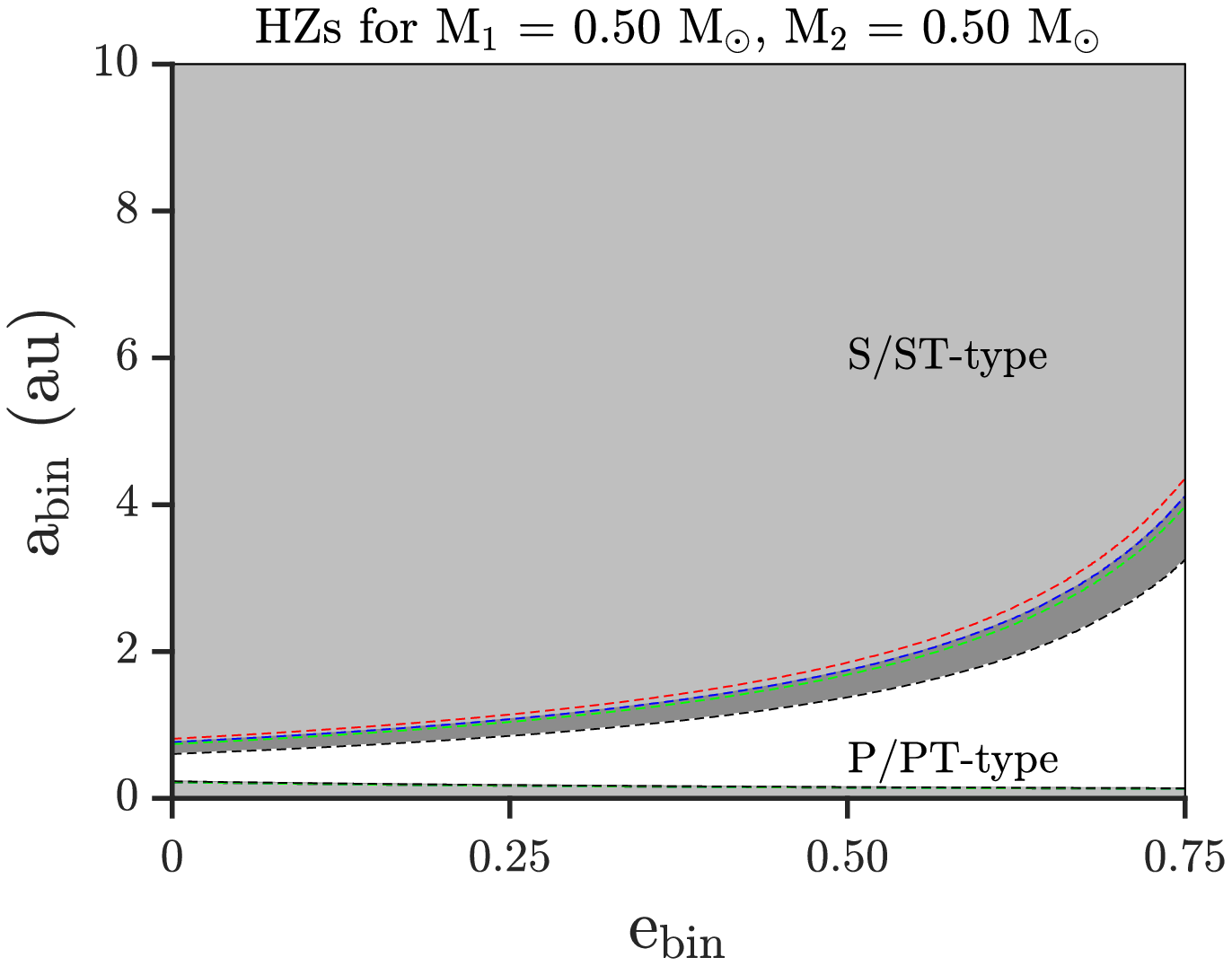} \\
\includegraphics[width=0.45\linewidth]{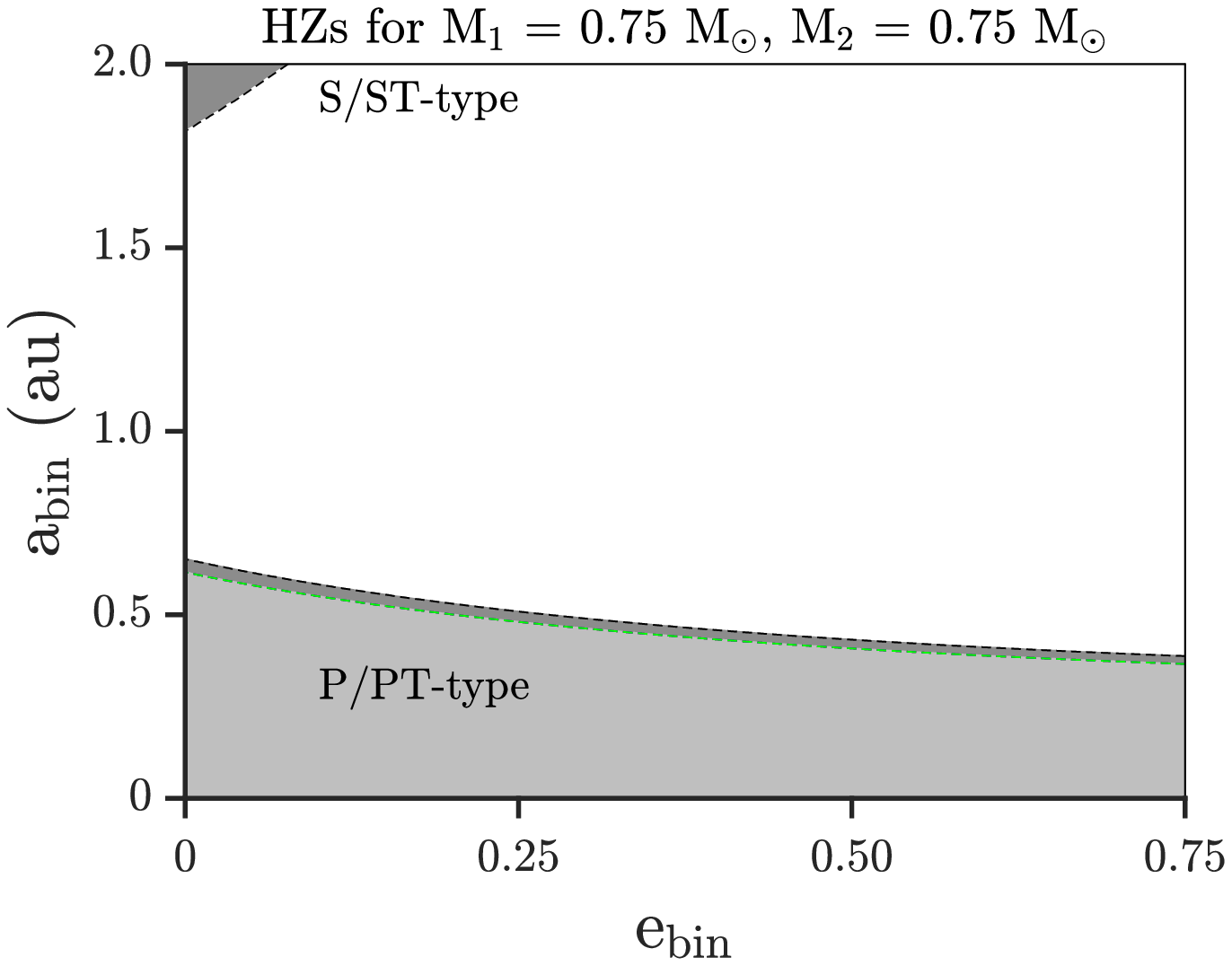} 
\hspace{0.5cm}
\includegraphics[width=0.45\linewidth]{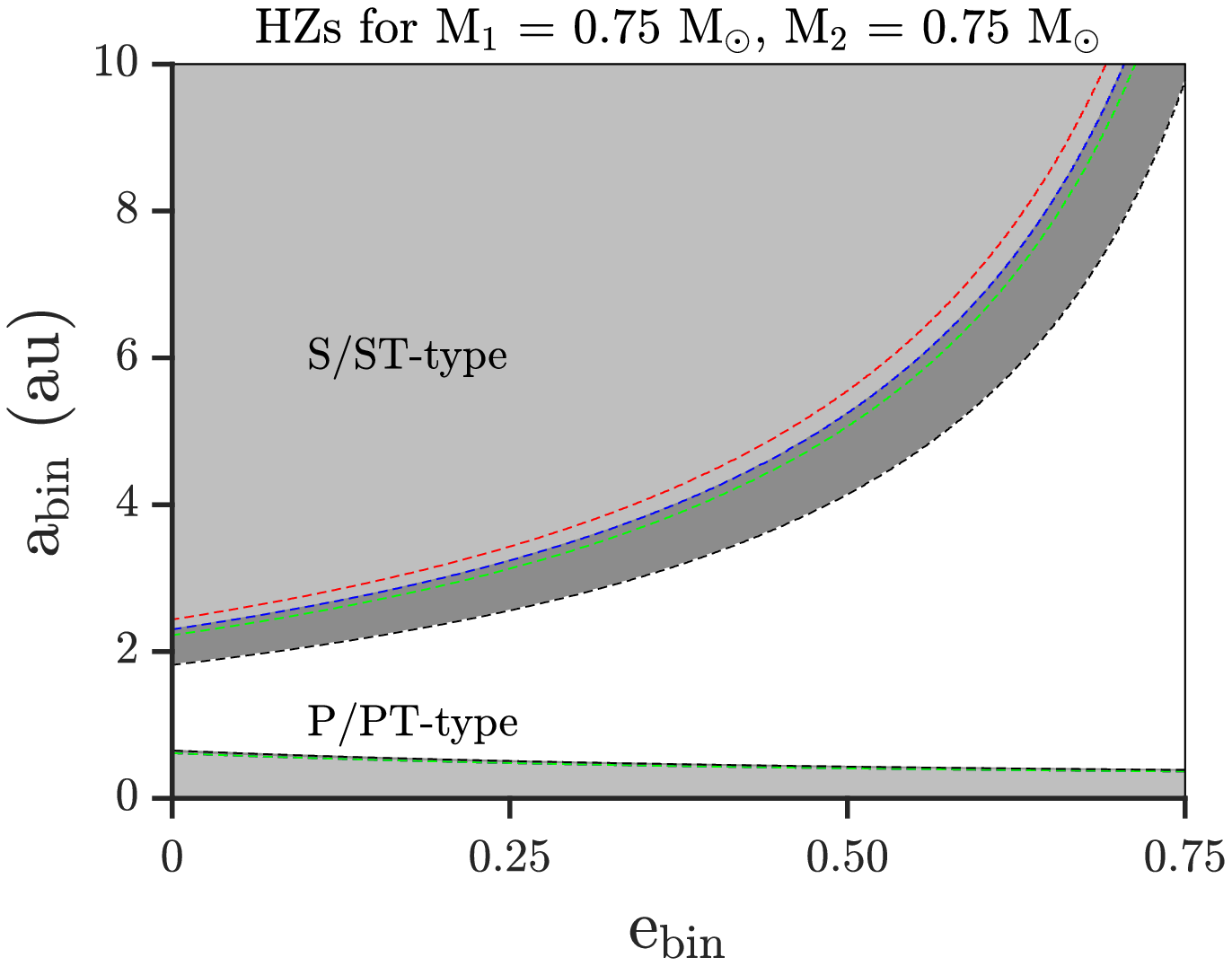} \\
\includegraphics[width=0.45\linewidth]{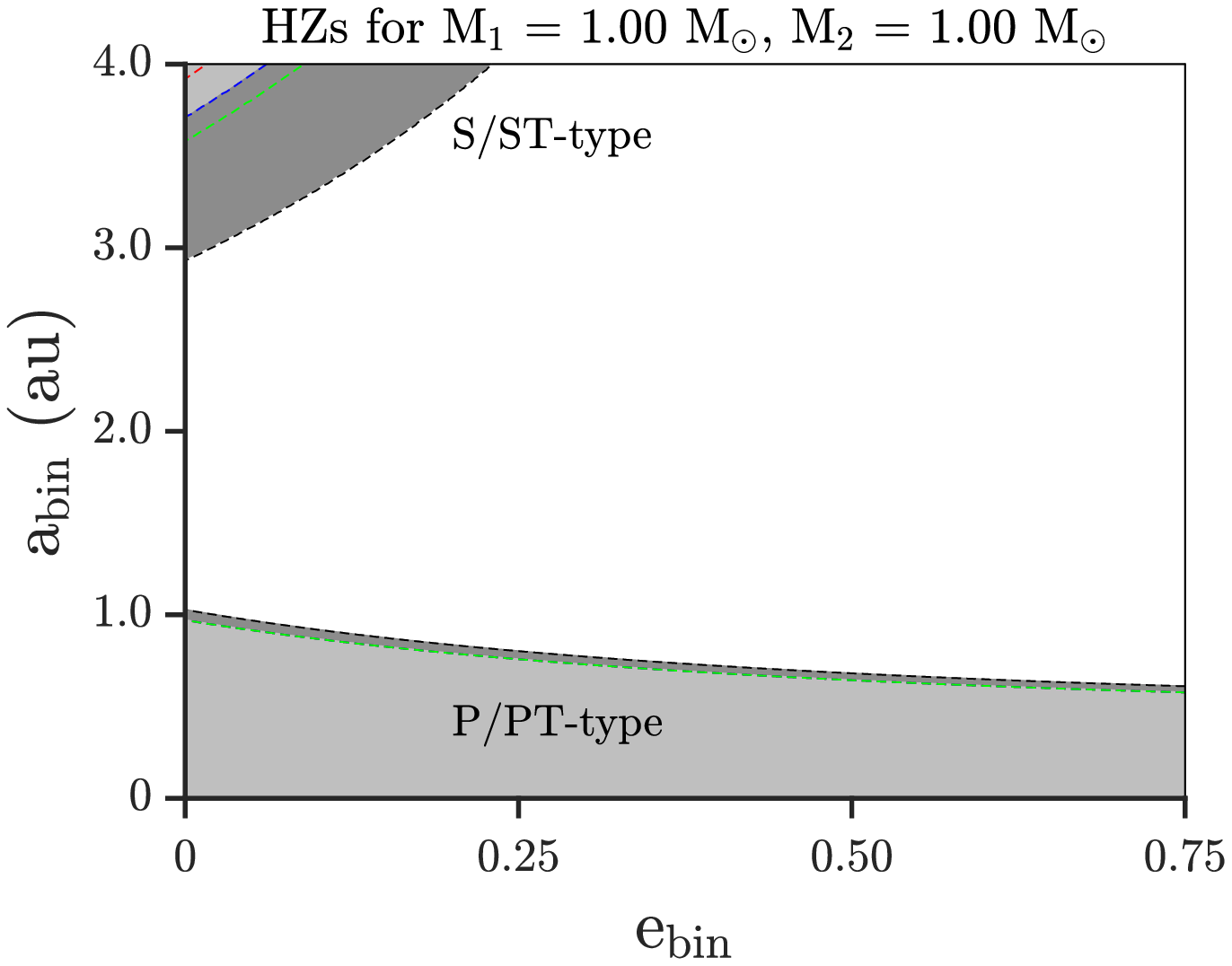} 
\hspace{0.5cm}
\includegraphics[width=0.45\linewidth]{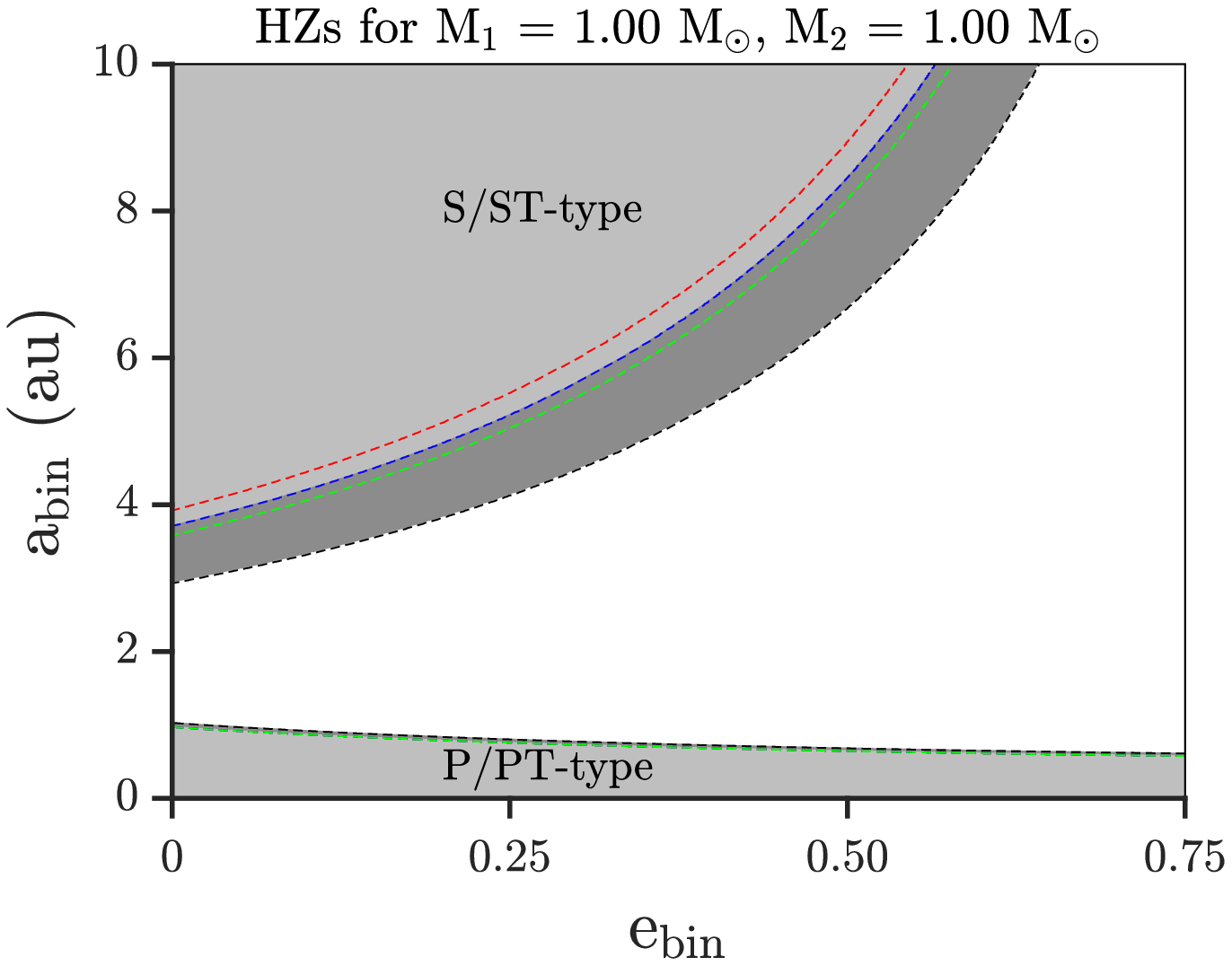} \\
\end{tabular}
\vspace{1cm}
\caption{
Case study 3: Calculated HZs and DZs for three equal-mass stellar binary systems where
$M_{1}$ = $M_{2}$ = 0.50, 0.75, and 1.00 $M_\mathrm{\odot}$.  The gray regions indicate 
the different possible HZs (i.e., S/ST-type and P/PT-type) with the white regions in between illustrating the 
DZs.  The dark and light gray regions correspond to the adopted definitions of GHZ and OHZ, respectively, 
assuming an Earth-mass planet; these definitions are also represented by black-dotted  lines (OHZ) and
blue-dotted lines (GHZE).  Additionally, the red-dotted and green-dotted lines illustrate the GHZ regions
assuming a Mars-mass (GHZM) and super-Earth-mass planet (GHZSE), respectively.
  }
\end{figure*}

%
%
\begin{figure*}
\centering
\begin{tabular} {c}
\includegraphics[width=0.45\linewidth]{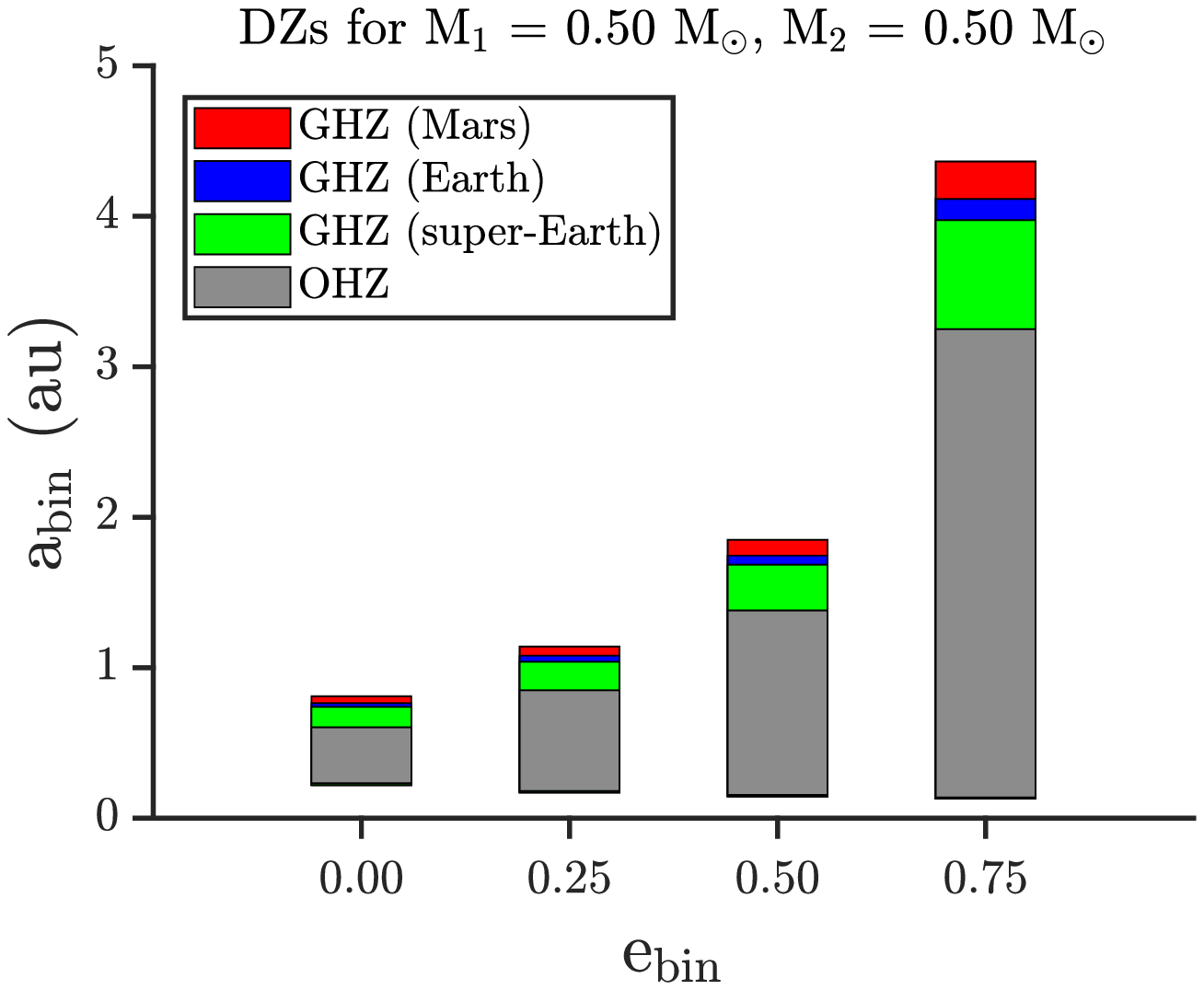} \\ 
\includegraphics[width=0.45\linewidth]{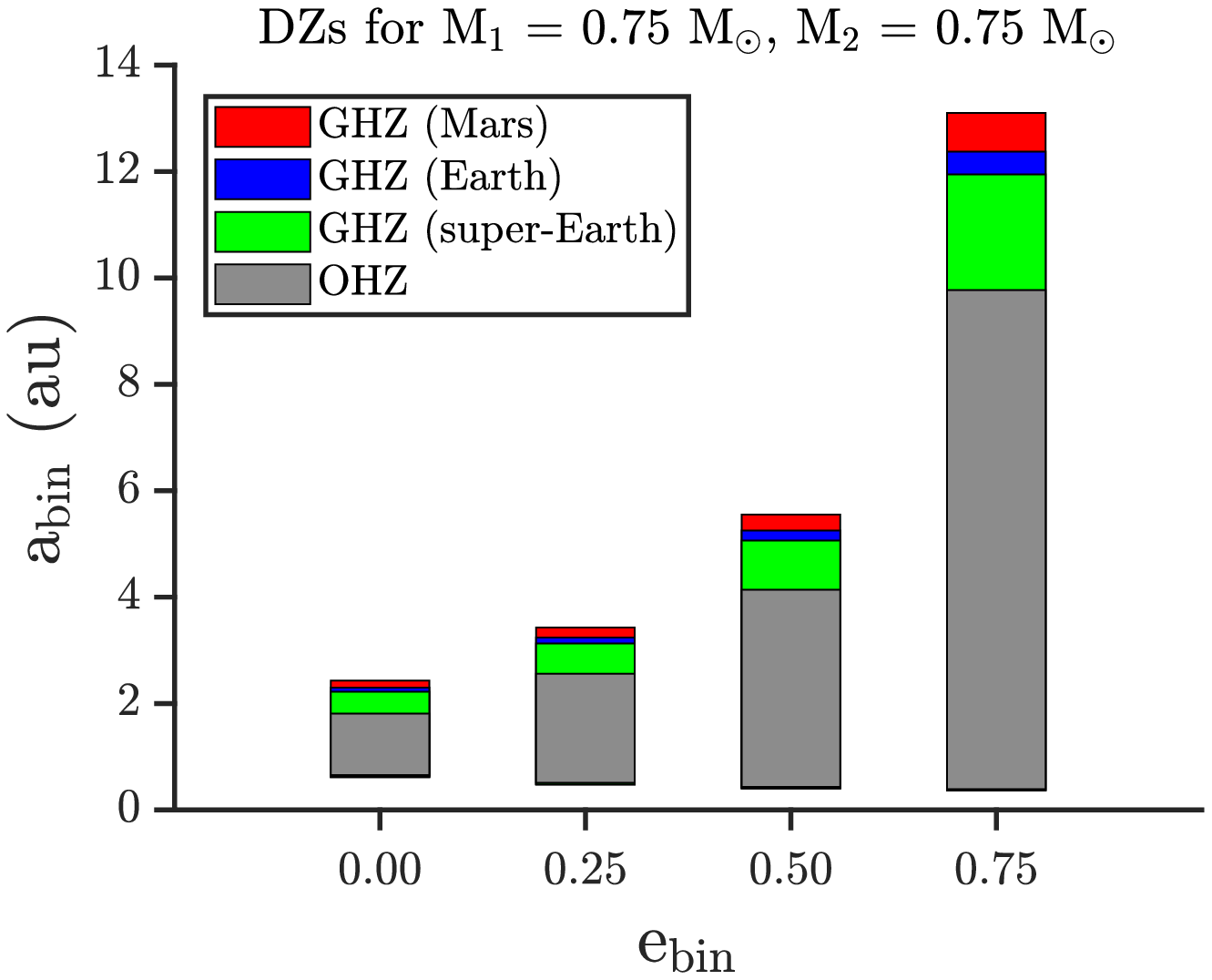} \\
\includegraphics[width=0.45\linewidth]{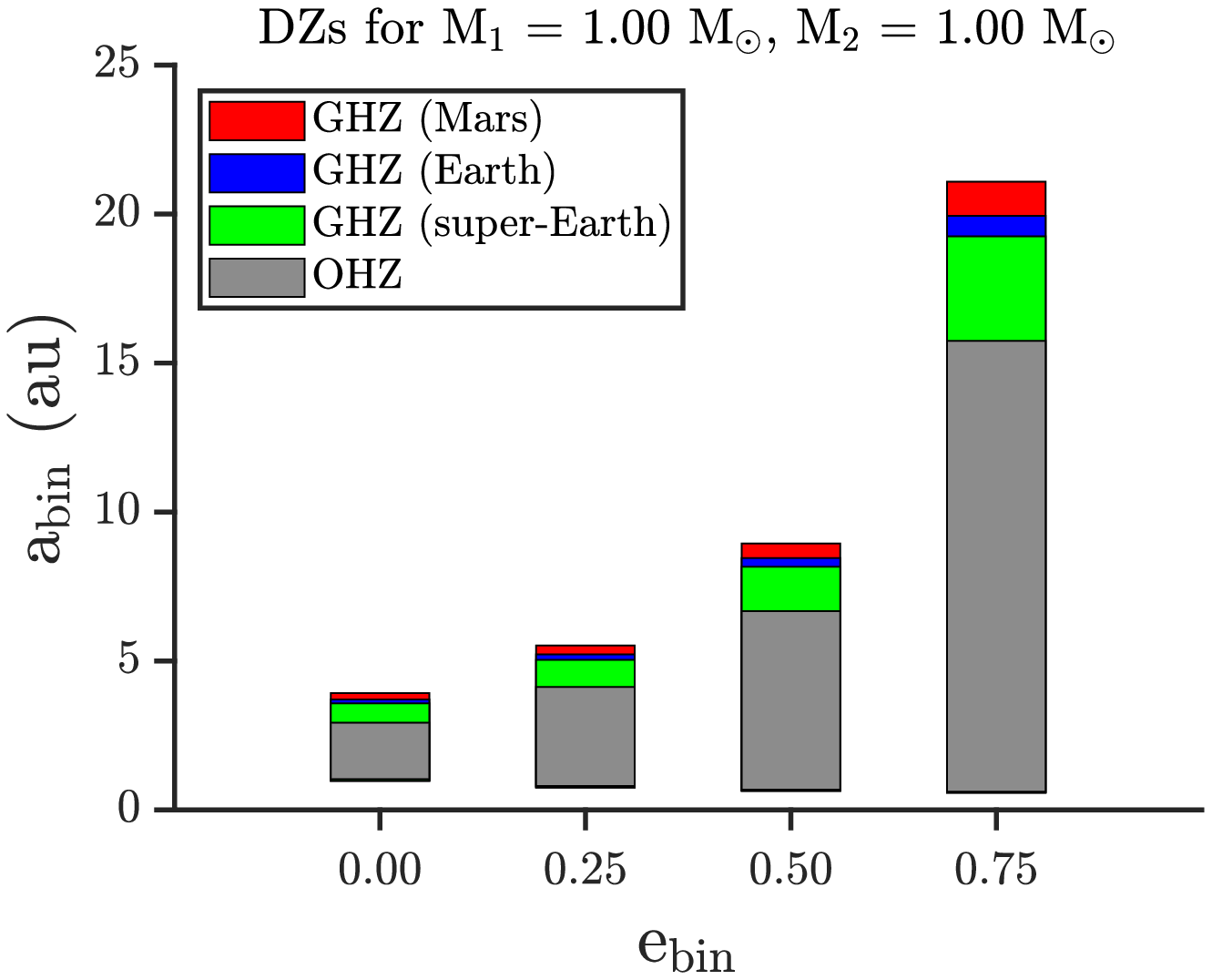} \\
\end{tabular}
\vspace{1cm}
\caption{
Case study 3: DZ widths for the three equal-mass stellar binary systems shown at selected binary 
eccentricities ($e_{\rm bin}$ = 0.00, 0.25, 0.50, and 0.75).  The gray region indicates the OHZ, whereas 
the red, blue, and green regions correspond to the GHZM, GHZE, and GHZSE case study, respectively.
  }
\end{figure*}

%
%
\begin{figure*}
\centering
\begin{tabular} {c}
\includegraphics[width=0.45\linewidth]{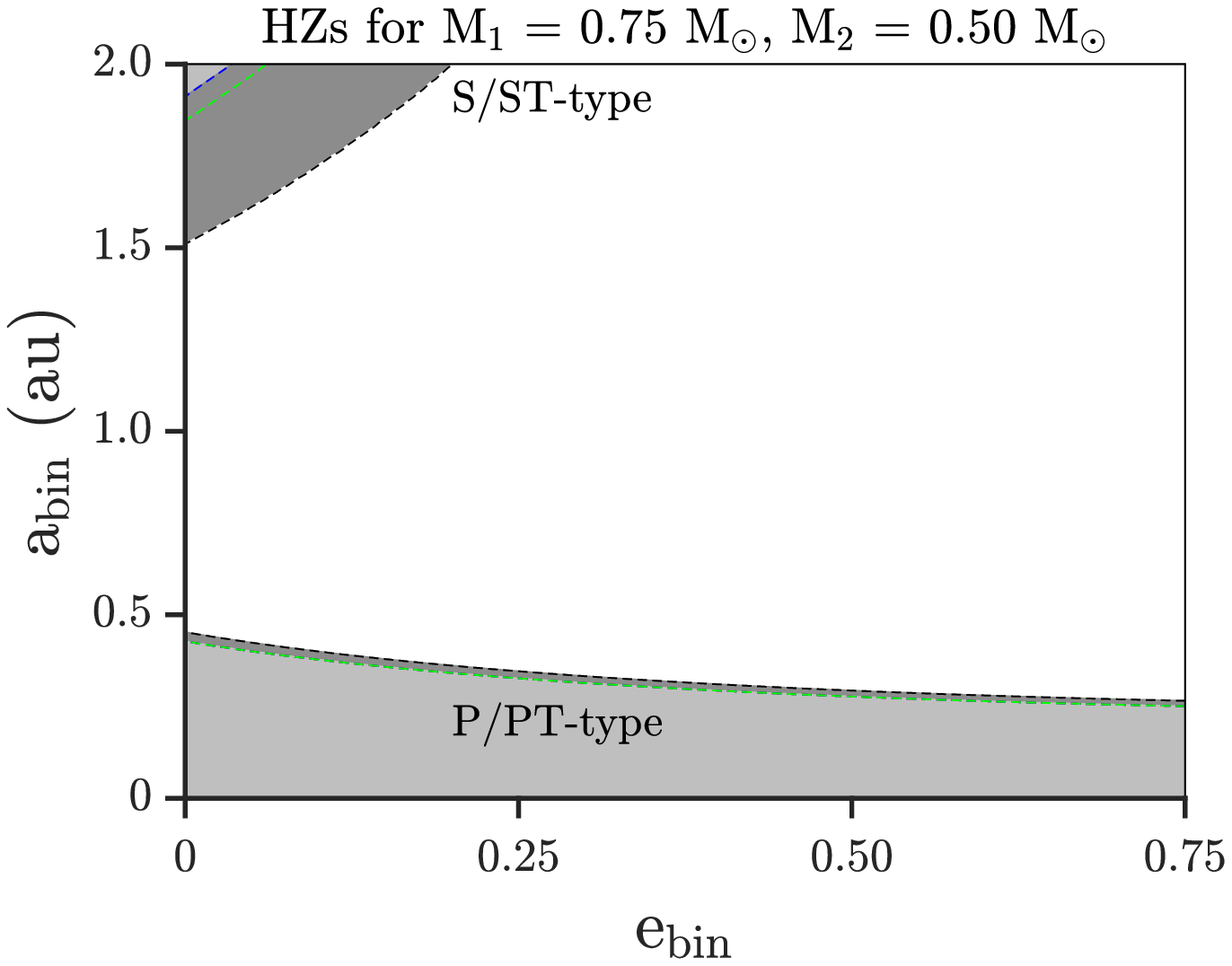} 
\hspace{0.5cm}
\includegraphics[width=0.45\linewidth]{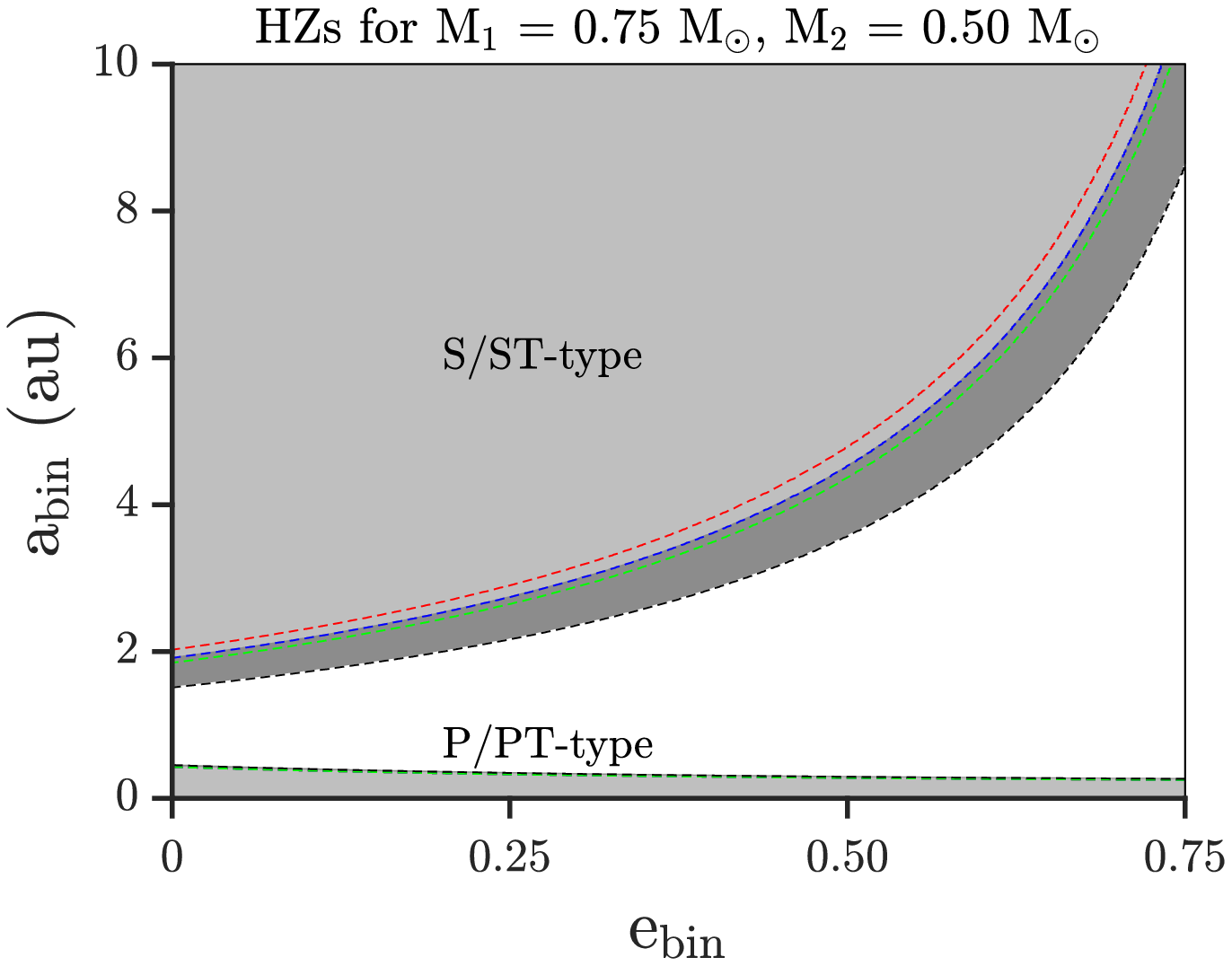} \\
\includegraphics[width=0.45\linewidth]{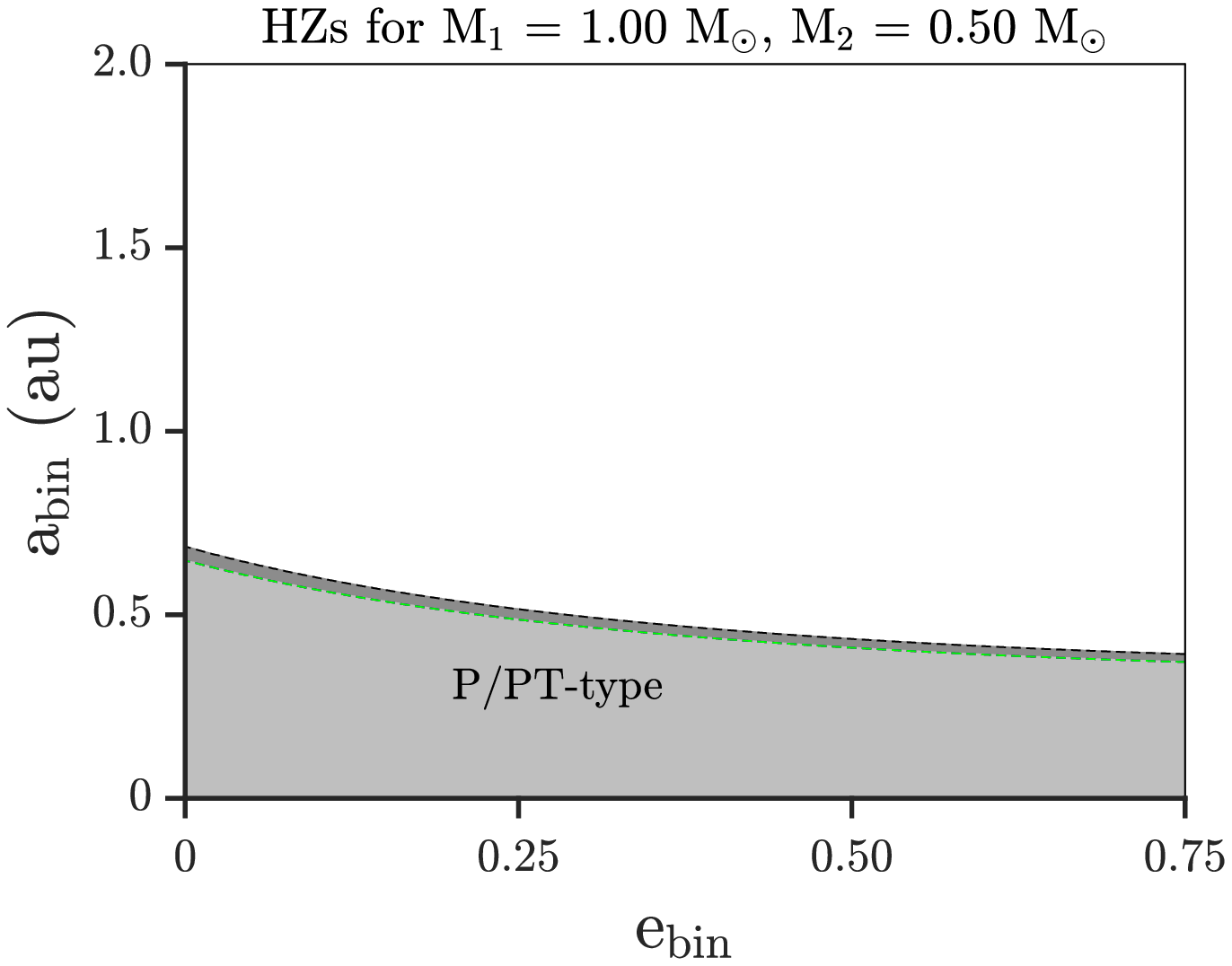}
\hspace{0.5cm}
\includegraphics[width=0.45\linewidth]{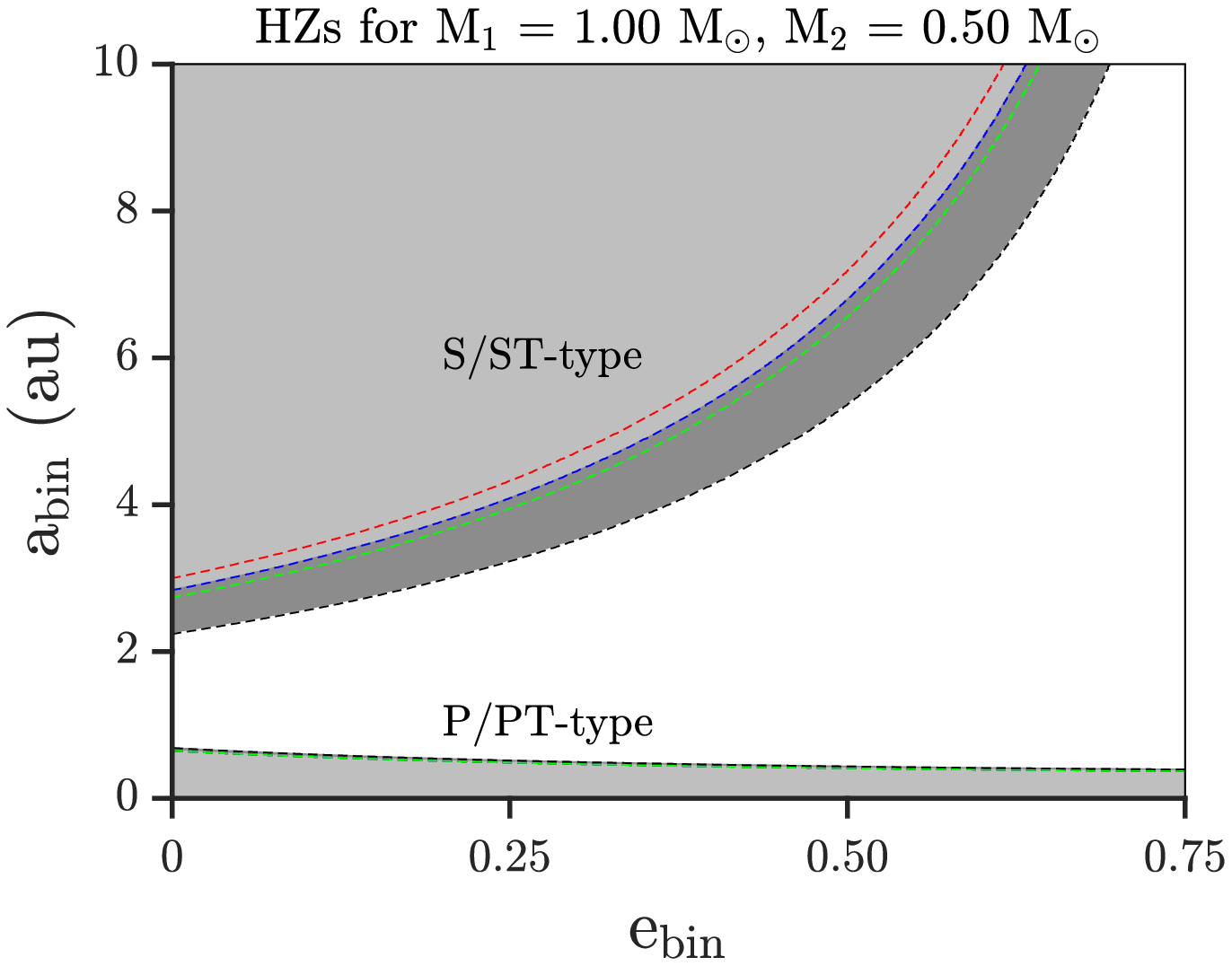} \\
\includegraphics[width=0.45\linewidth]{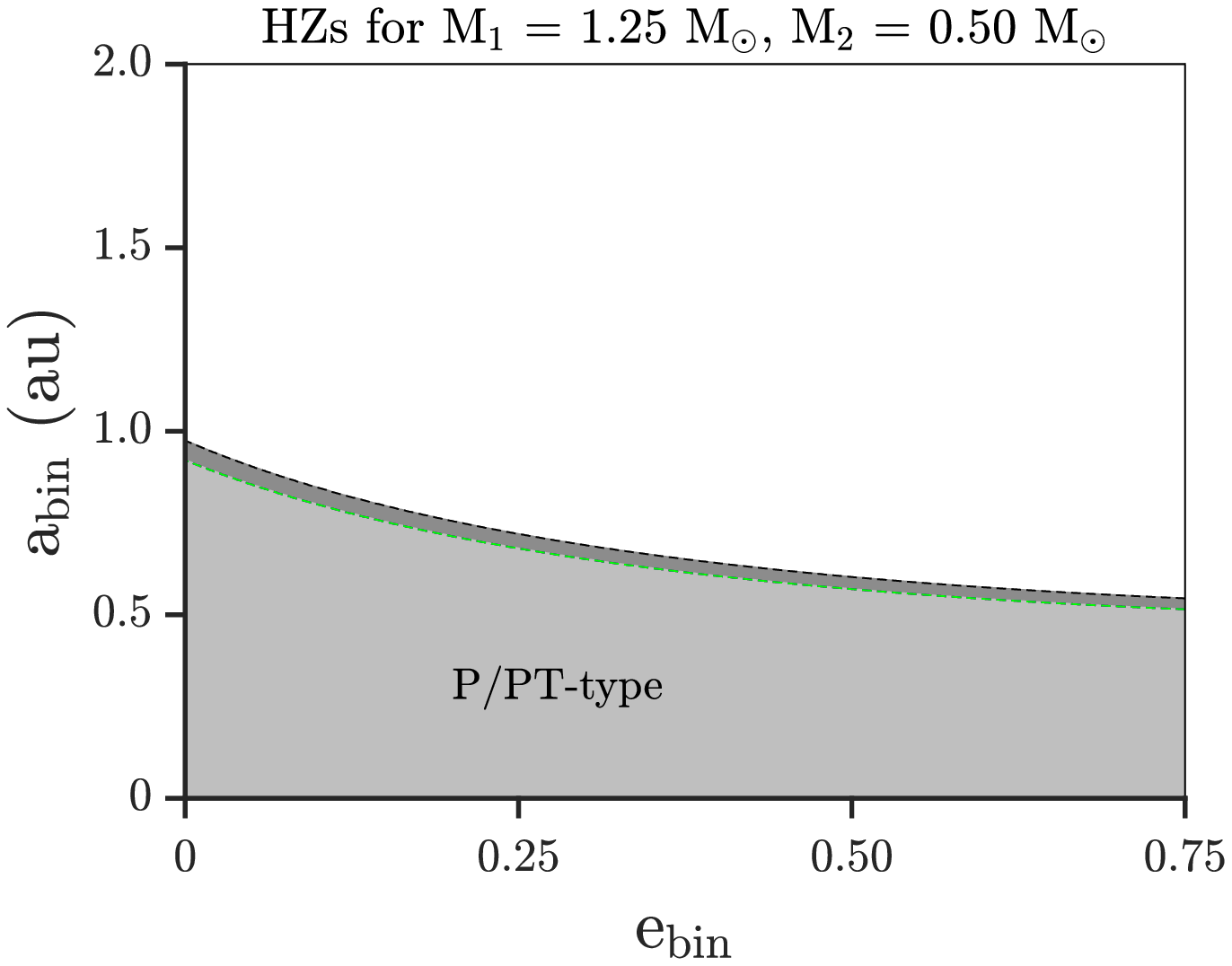}
\hspace{0.5cm}
\includegraphics[width=0.45\linewidth]{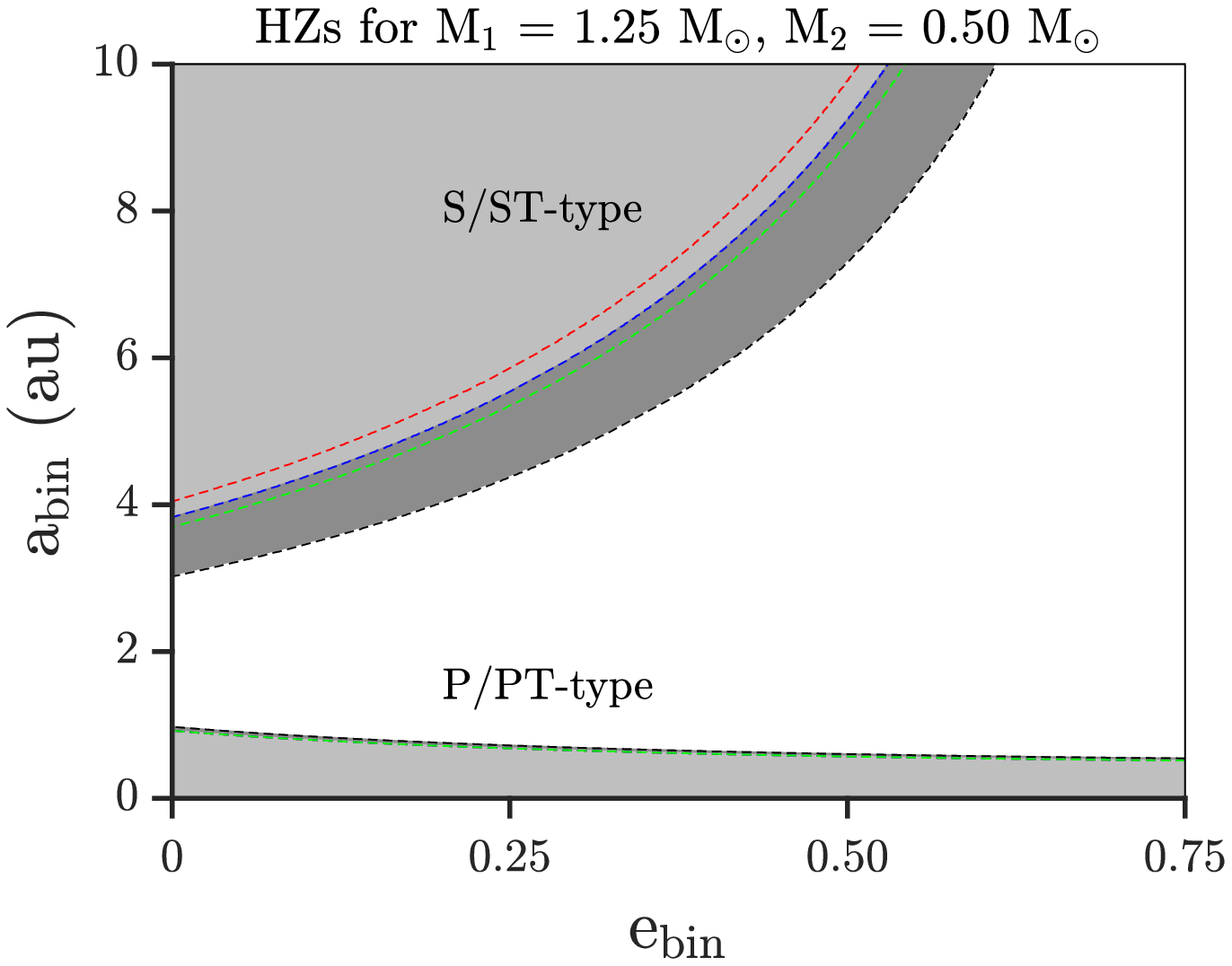} \\
\end{tabular}
\vspace{1cm}
\caption{
Case study 3: Same as Fig.~5, but for three nonequal-mass stellar binary systems where 
$M_{1}$ = 0.75, 1.00, and 1.25 $M_\mathrm{\odot}$ and $M_{2}$ = 0.50 $M_\mathrm{\odot}$.
}
\end{figure*}

%
%
\begin{figure*}
\centering
\begin{tabular} {c}
\includegraphics[width=0.5\linewidth]{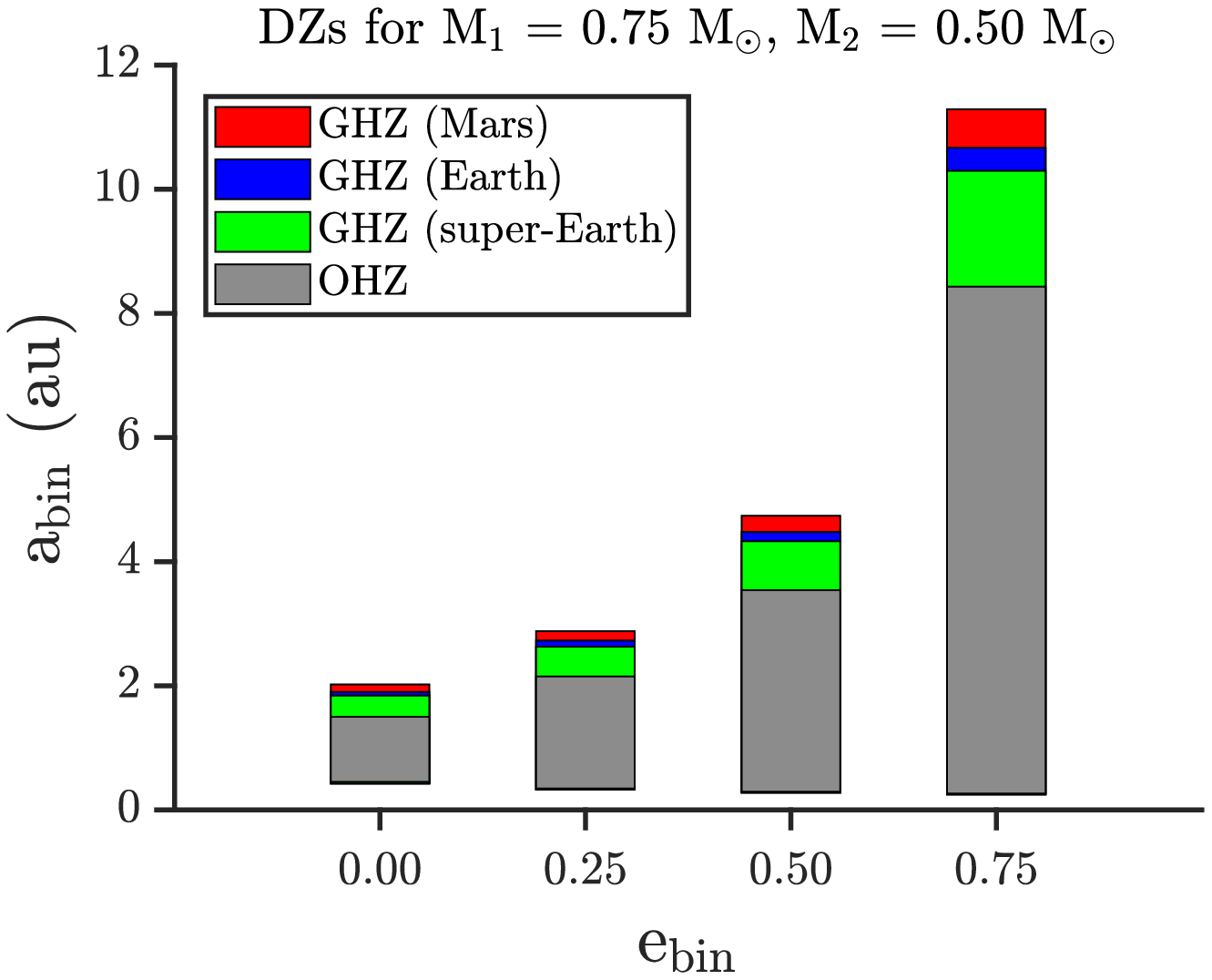} \\
\includegraphics[width=0.5\linewidth]{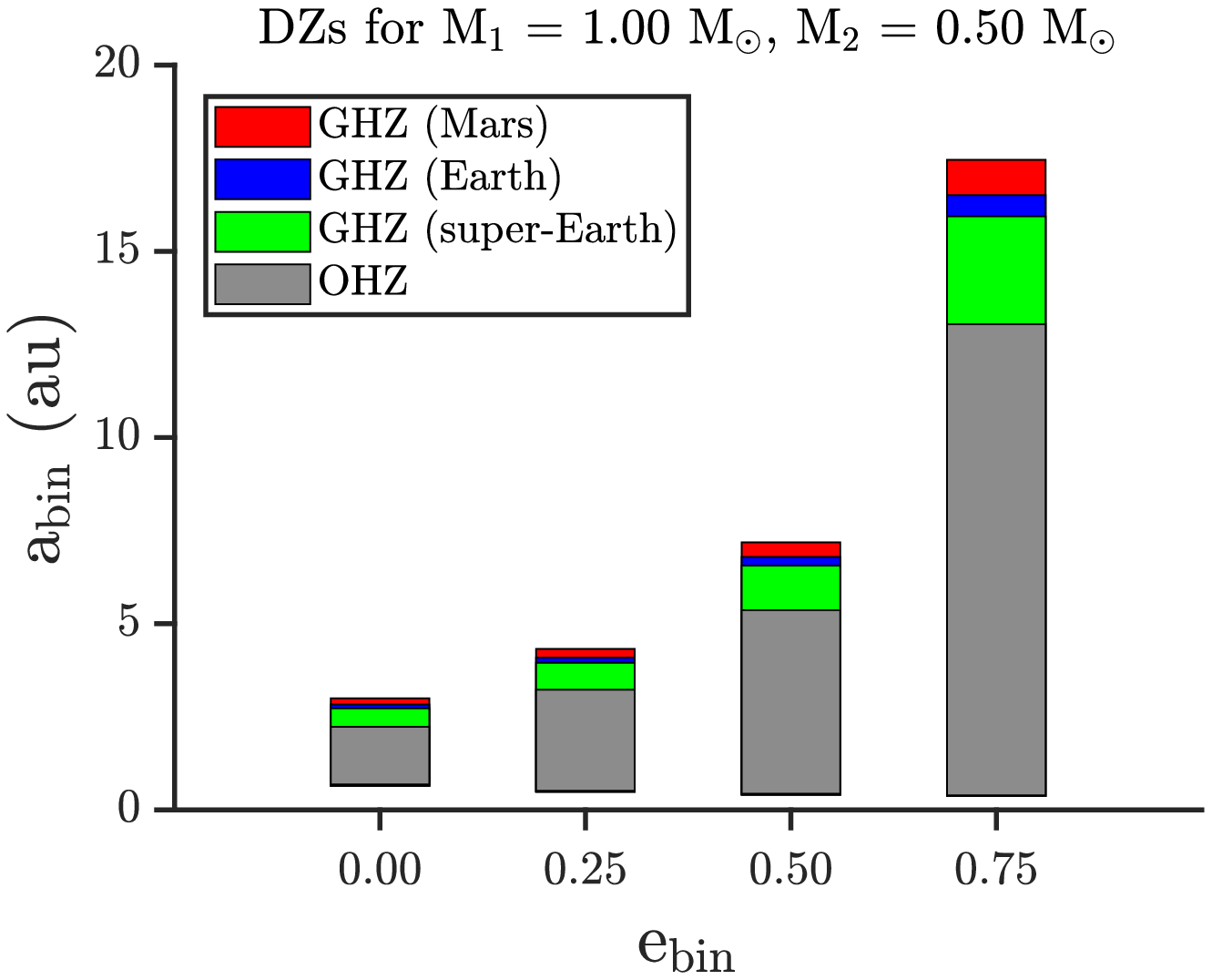} \\
\includegraphics[width=0.5\linewidth]{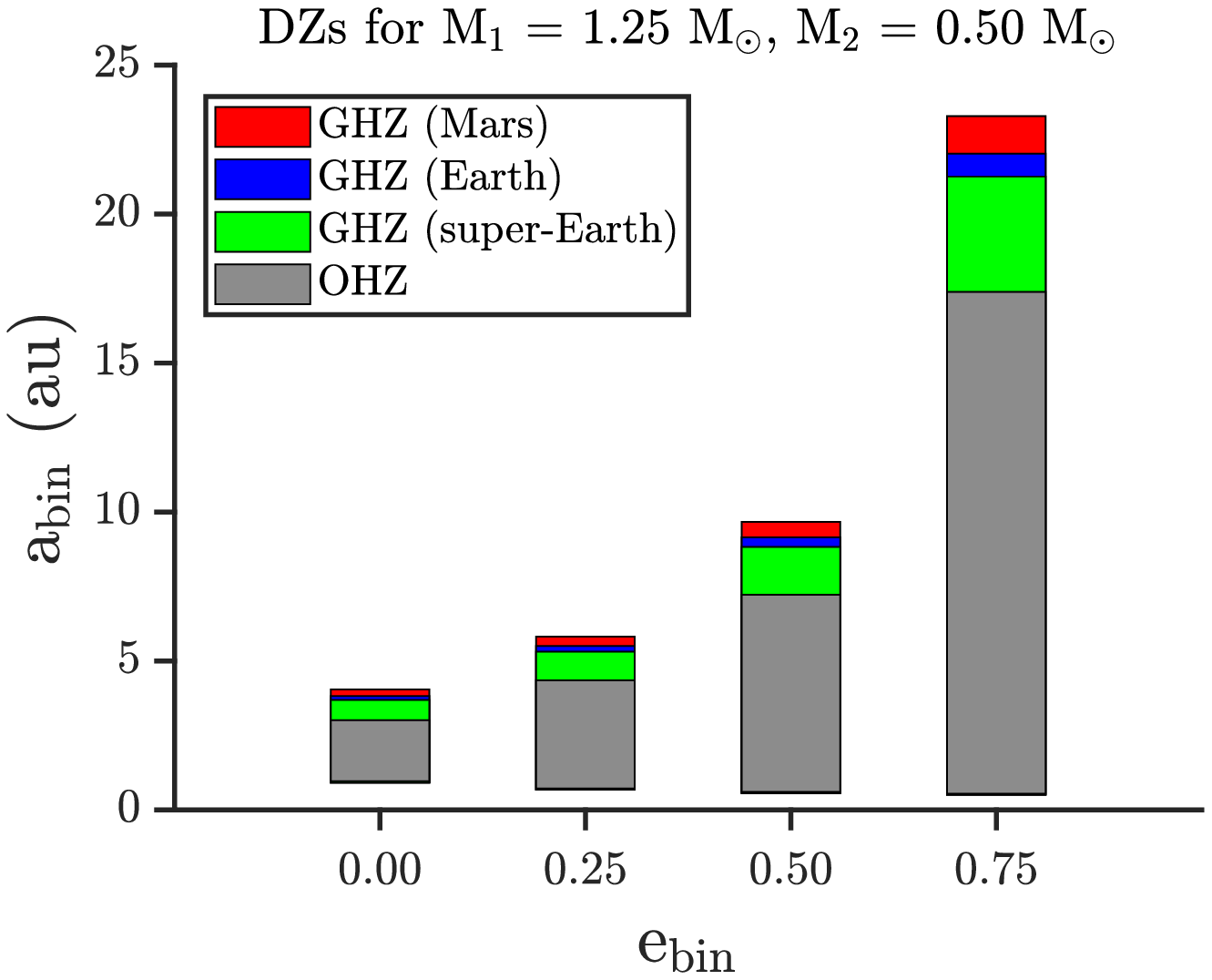} \\
\end{tabular}
\vspace{1cm}
\caption{
Case study 3: Same as Fig.~6, but for three nonequal-mass stellar binary systems where 
$M_{1}$ = 0.75, 1.00, and 1.25 $M_\mathrm{\odot}$ and $M_{2}$ = 0.50 $M_\mathrm{\odot}$.
}
\end{figure*}

%
%
\begin{figure*}
\centering
\begin{tabular} {c}
\includegraphics[width=0.45\linewidth]{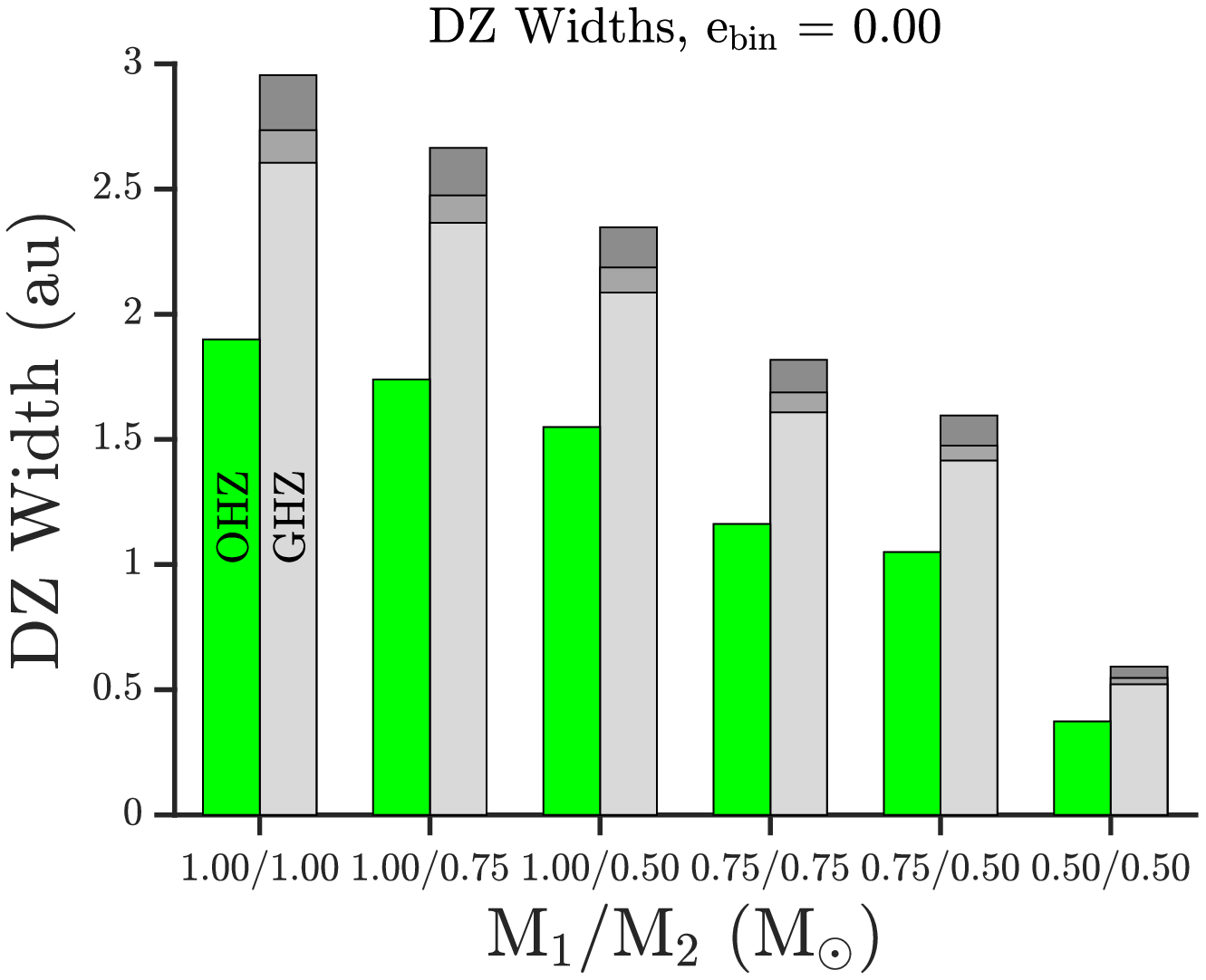} 
\hspace{0.5cm}
\includegraphics[width=0.45\linewidth]{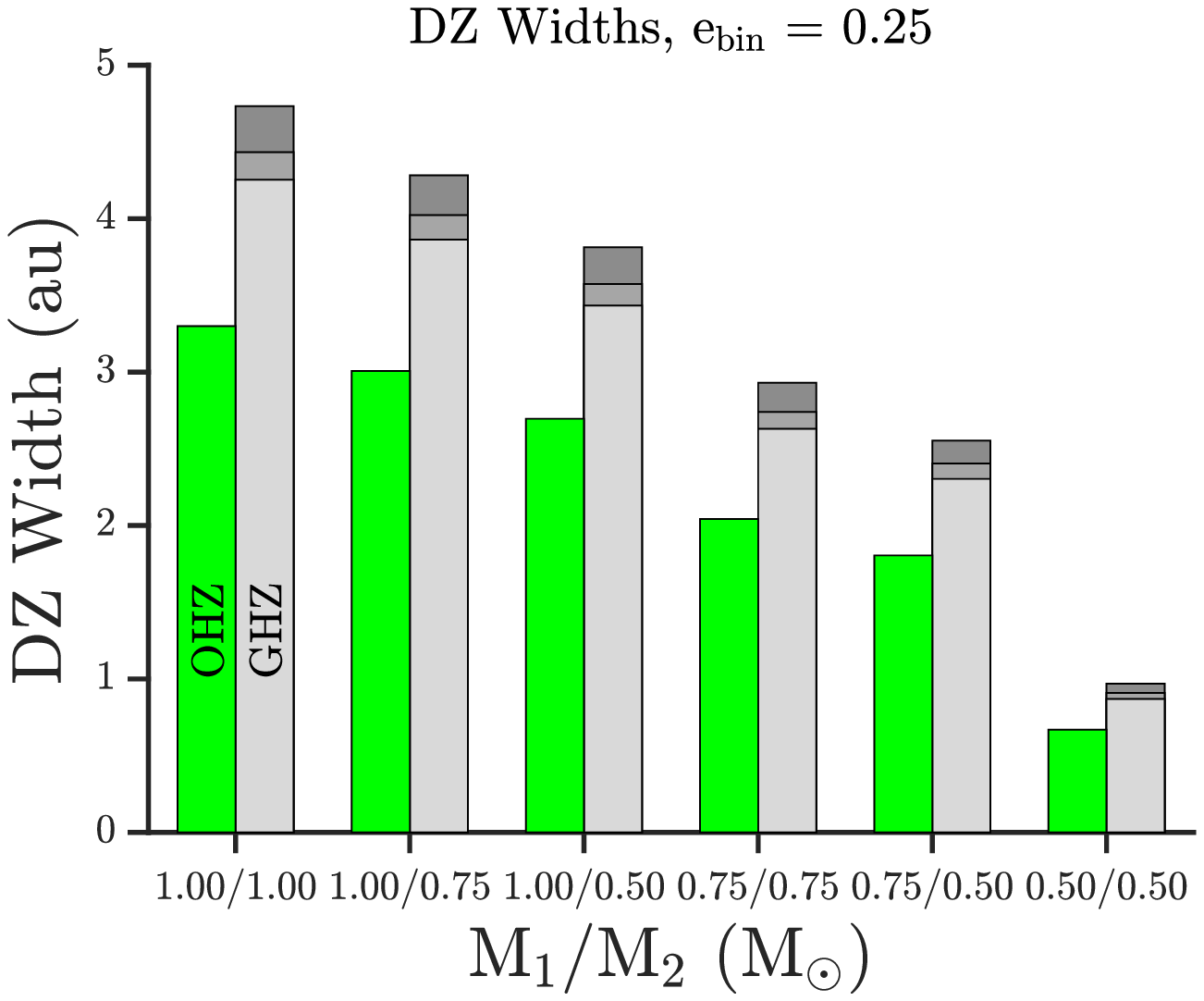} \\ \\
\includegraphics[width=0.45\linewidth]{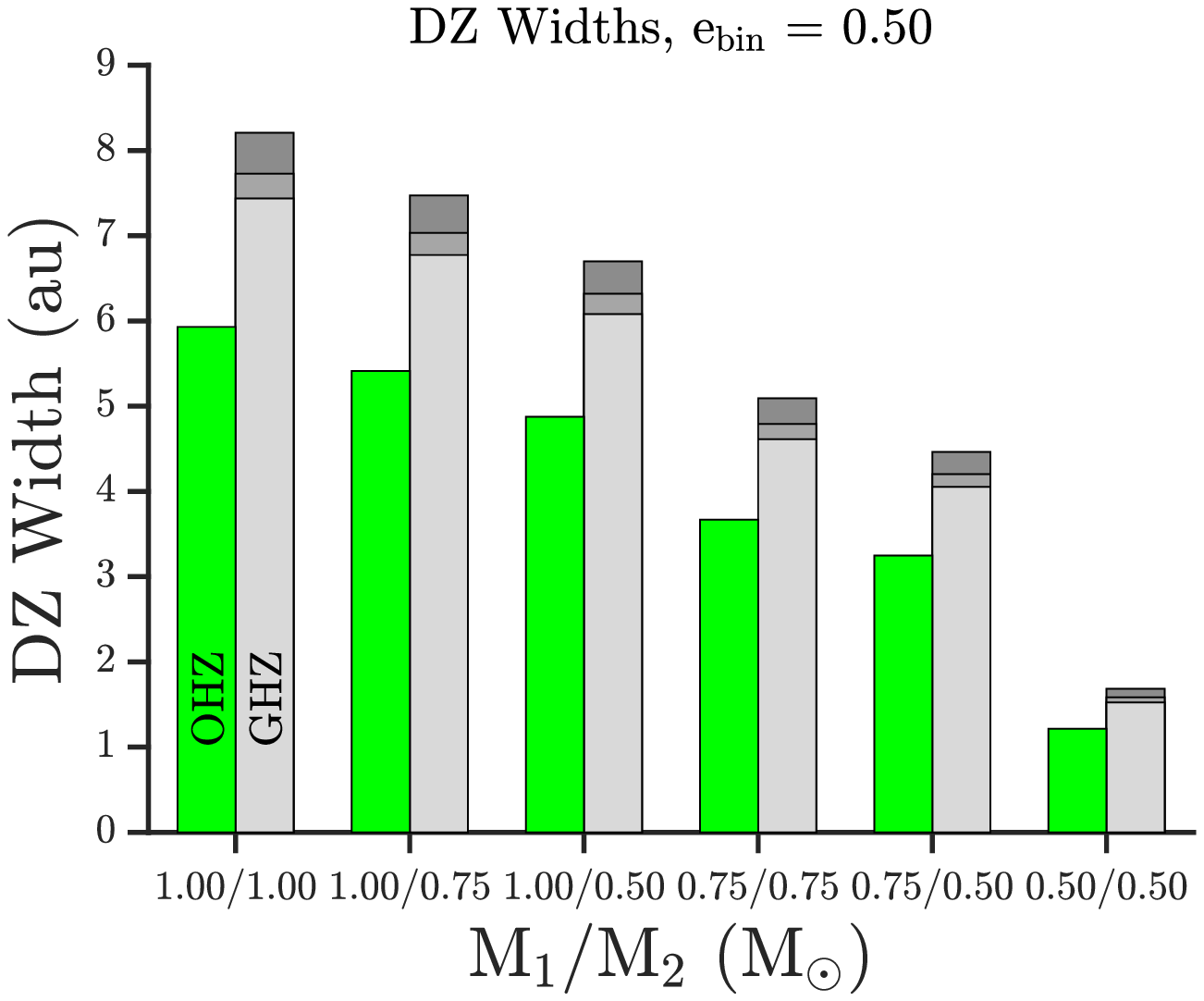} 
\hspace{0.5cm}
\includegraphics[width=0.45\linewidth]{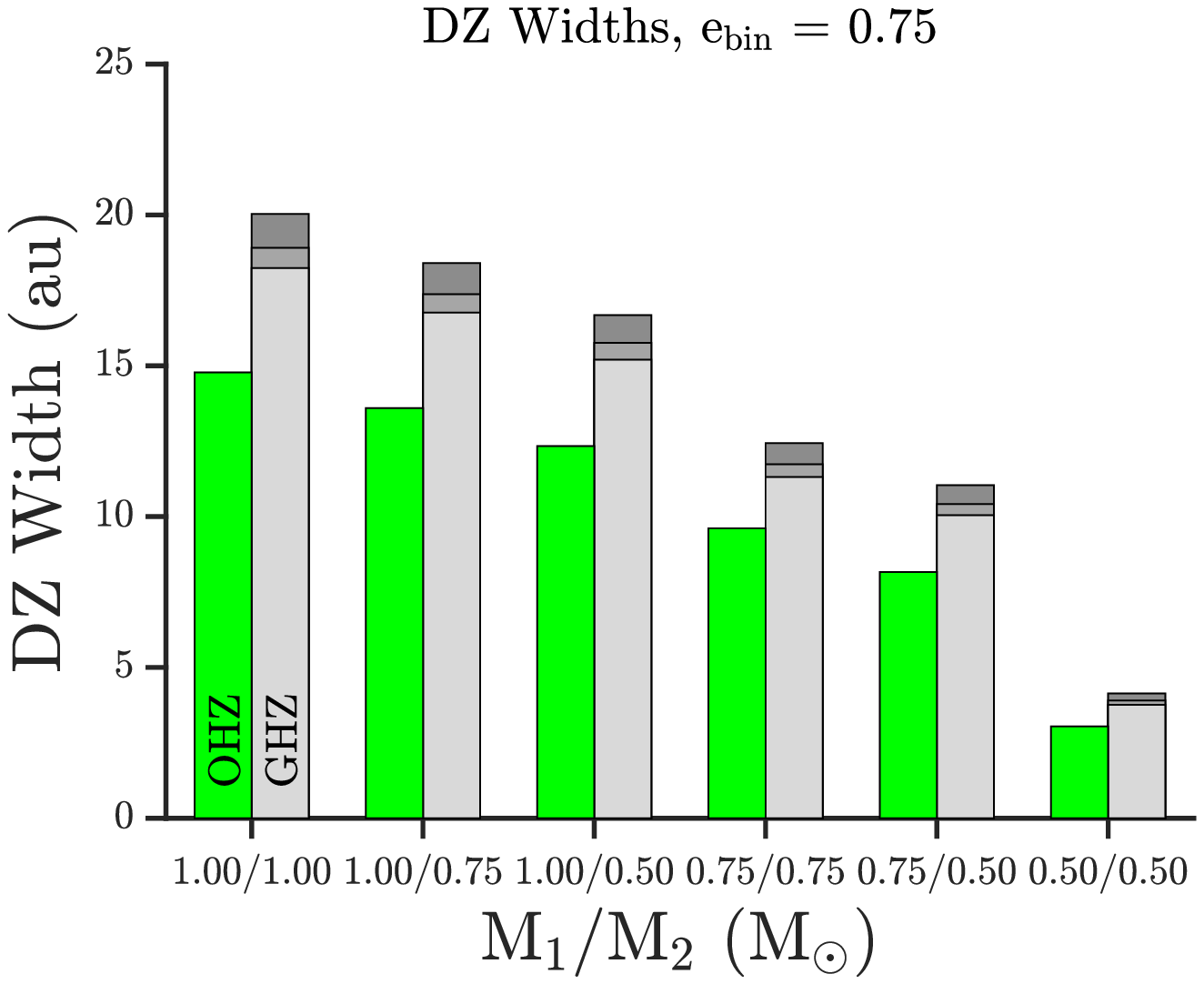} \\
\end{tabular}
\vspace{1cm}
\caption{
Case study 3: DZ widths for several stellar binary systems at selected binary eccentricities 
($e_{\rm bin}$ = 0.00, 0.25, 0.50, and 0.75).  The gray regions indicate the different 
DZs corresponding to the GHZ criterion with light, medium, and dark gray corresponding
to super-Earth-mass, Earth-mass, and Mars-mass planets, respectively.  The green regions represent
the OHZ criterion; following \cite{kop14} (see text), it does not depend on the planetary mass.
The horizontal axes convey the stellar binary systems in terms of primary and secondary mass
($M_{1}/M_{2}$) with the vertical axes illustrating the binary semi-major axes, $a_{\mathrm{bin}}$.
}
\end{figure*}

\end{document}